\documentclass[10pt, twocolumn, amsmath, amssymb]{quantumarticle}

\usepackage{tikz}
\usepackage[strict,uselistings]{revquantum}
	\newrm{ir}
	\newrm{ctrl}
	\newnew{cal}{\mathcal}
	\newcal{GP}

	\newaffil{UTSQSI}{
		University of Technology Sydney,
		Centre for Quantum Software and Information,
		Ultimo NSW 2007, Australia
	 }

	\newaffil{CQDGriffith}{
		Centre for Quantum Dynamics,
		Griffith University,  Brisbane, Queensland  4111,  Australia
	}
	\newaffil{CQCCTGriffith}{
		Centre for Quantum Computation and Communication Technology,
		Griffith University,  Brisbane, Queensland  4111,  Australia
	}

    \newcommand{\apx}[1]{\hyperref[apx:#1]{Appendix~\ref*{apx:#1}}}

\usepackage[bold]{hhtensor}

\usepackage[sort&compress,numbers,merge]{natbib}

\algnewcommand{\LineComment}[1]{\State \(\triangleright\) #1}

\algblockdefx[Switch]{Switch}{EndSwitch}[1]{\textbf{switch} #1}[0]{\textbf{end switch}}
\algloopdefx[Case]{Case}[1]{\textbf{case} #1}

\usepackage[caption=false]{subfig}

\makeatletter
\def\Decl@Mn@Delim#1#2#3#4{%
  \if\relax\noexpand#1%
    \let#1\undefined
  \fi
  \DeclareMathDelimiter{#1}{#2}{#3}{#4}{#3}{#4}}
\def\Decl@Mn@Open#1#2#3{\Decl@Mn@Delim{#1}{\mathopen}{#2}{#3}}
\def\Decl@Mn@Close#1#2#3{\Decl@Mn@Delim{#1}{\mathclose}{#2}{#3}}
\DeclareFontFamily{OMX}{MnSymbolE}{}
\DeclareFontShape{OMX}{MnSymbolE}{m}{n}{
    <-6>  MnSymbolE5
   <6-7>  MnSymbolE6
   <7-8>  MnSymbolE7
   <8-9>  MnSymbolE8
   <9-10> MnSymbolE9
  <10-12> MnSymbolE10
  <12->   MnSymbolE12}{}
\DeclareFontShape{OMX}{MnSymbolE}{b}{n}{
    <-6>  MnSymbolE-Bold5
   <6-7>  MnSymbolE-Bold6
   <7-8>  MnSymbolE-Bold7
   <8-9>  MnSymbolE-Bold8
   <9-10> MnSymbolE-Bold9
  <10-12> MnSymbolE-Bold10
  <12->   MnSymbolE-Bold12}{}
\DeclareSymbolFont{mnsymbols}  {OMX}{MnSymbolE}{m}{n}
\Decl@Mn@Open {\lsem}               {mnsymbols}{'102}
\Decl@Mn@Close{\rsem}               {mnsymbols}{'107}
\Decl@Mn@Open {\llangle}            {mnsymbols}{'164}
\Decl@Mn@Close{\rrangle}            {mnsymbols}{'171}
\makeatother

\newcommand{\figurefolder}{../fig}

\renewcommand{\figurefolder}{fig}
\begin{document}

%=============================================================================
% FRONT MATTER
%=============================================================================

\title{Bayesian Quantum Noise Spectroscopy}
\date{\today}

\author{Chris Ferrie}
	\affilUTSQSI

\author{Chris Granade}
	\affilEQuSUSyd
	\affilUSydPhys

\author{Gerardo Paz-Silva}
	\affilCQDGriffith
	\affilCQCCTGriffith
	
\author{Howard Wiseman}
	\affilCQDGriffith
	\affilCQCCTGriffith

\begin{abstract}
	As commonly understood, the noise spectroscopy problem---characterizing the statistical properties of a noise process affecting a quantum system by measuring its response---is ill-posed. Ad-hoc solutions assume implicit structure which is often never determined. Thus it is unclear when the method will succeed or whether one should trust the solution obtained. Here we propose to treat the problem from the point of view of statistical estimation theory. We develop a Bayesian solution to the problem which allows one to easily incorporate assumptions which render the problem solvable. We compare several numerical techniques for noise spectroscopy and find the Bayesian approach to be superior in many respects. 
\end{abstract}

\maketitle

\tableofcontents

%=============================================================================
\section{Introduction}
\label{sec:intro}
%=============================================================================

Quantum technologies will very likely ultimately rely on active error correction. However, at every stage---crucially in current experiments---open-loop control techniques to \emph{suppress} errors need to be employed~\cite{QECBook}. They can be thought of as a layer-0 level of protection designed to make the errors in any operation as small as possible, before the machinery of quantum error correction and fault-tolerant quantum computing takes over. 

These techniques can be roughly classified in terms of their robustness to uncertainty in the knowledge of the noise they attempt to suppress. Exemplifying this, dynamical decoupling (DD)~\cite{Viola98} sits on one end of the spectrum as a highly robust technique---DD sequences only require the noise to be `slow' (in some appropriate metric we will discuss in more detail later), but beyond this the details of the noise are not important. On the other hand, optimal control techniques can be used to design pulse sequences capable of efficiently suppressing a wide range of noises, both `fast' and `slow', but only if detailed knowledge of the noise is available~\cite{OptCtrl, Calarco}. The ideal strategy is thus a function of the knowledge available about the noise. Unfortunately this detailed information is often absent, since in an open quantum system scenario the bath or environment generating such noise cannot be directly measured or controlled. Phenomenological models and intimate control over the fabrication process of a given quantum system can alleviate this, but can at most give partial information about the noise sources affecting a qubit. 

In order to bridge this gap in knowledge, {\it quantum noise spectroscopy} protocols of varying generality have been developed and implemented in recent years~\cite{yan_spectroscopy_2012, yan_rotating-frame_2013, dial_charge_2013, yuge_measurement_2011, young_qubits_2012, Alvarez2011, Norris2016, paz-silva_multiqubit_2017, Muhonen2014}. Their objective is to characterize the actual noise affecting a quantum system of interest, regardless of its source, in terms of its correlations, or more specifically the set of power poly-spectra~\cite{Breuer:book}. The key point is that the information these protocols output, together with optimal control techniques, should enable one to design control routines tailored to suppress the actual noise affecting the quantum system of interest \cite{shulman_suppressing_2014}. Very recently, for example, a 10-minute record-breaking coherence time was achieved in trapped-ions~\cite{Wang2017} using this principle. Operationally, spectroscopy protocols measure the response of a quantum system, in terms of expectation values of observables, in a known initial state, to the noise affecting it in tandem with user-determined control routines. The main difficulty is that noise correlations influence the dynamics of the quantum system in a highly non-linear way. Thus, inferring these correlations in detail from the response of the quantum system is generally an ill-posed problem, unless a priori information on the noise is assumed. Even when standard assumptions, such as Gaussian noise or a dephasing coupling are satisfied, the problem remains non-linear and inverting it carries along a set of non-trivial complications that in turn constrain the type of noise that can be characterized. For example, in Refs.~\cite{Alvarez2011, yuge_measurement_2011, young_qubits_2012, Norris2016, paz-silva_multiqubit_2017} a control-induced frequency comb approach is used in order to overcome the non-linear character of the problem but it comes at the cost of being only effective when the noise correlations are {\it smooth} functions in frequency space.   

\begin{figure}
	\begin{center}
		\includegraphics[width=0.95\columnwidth]{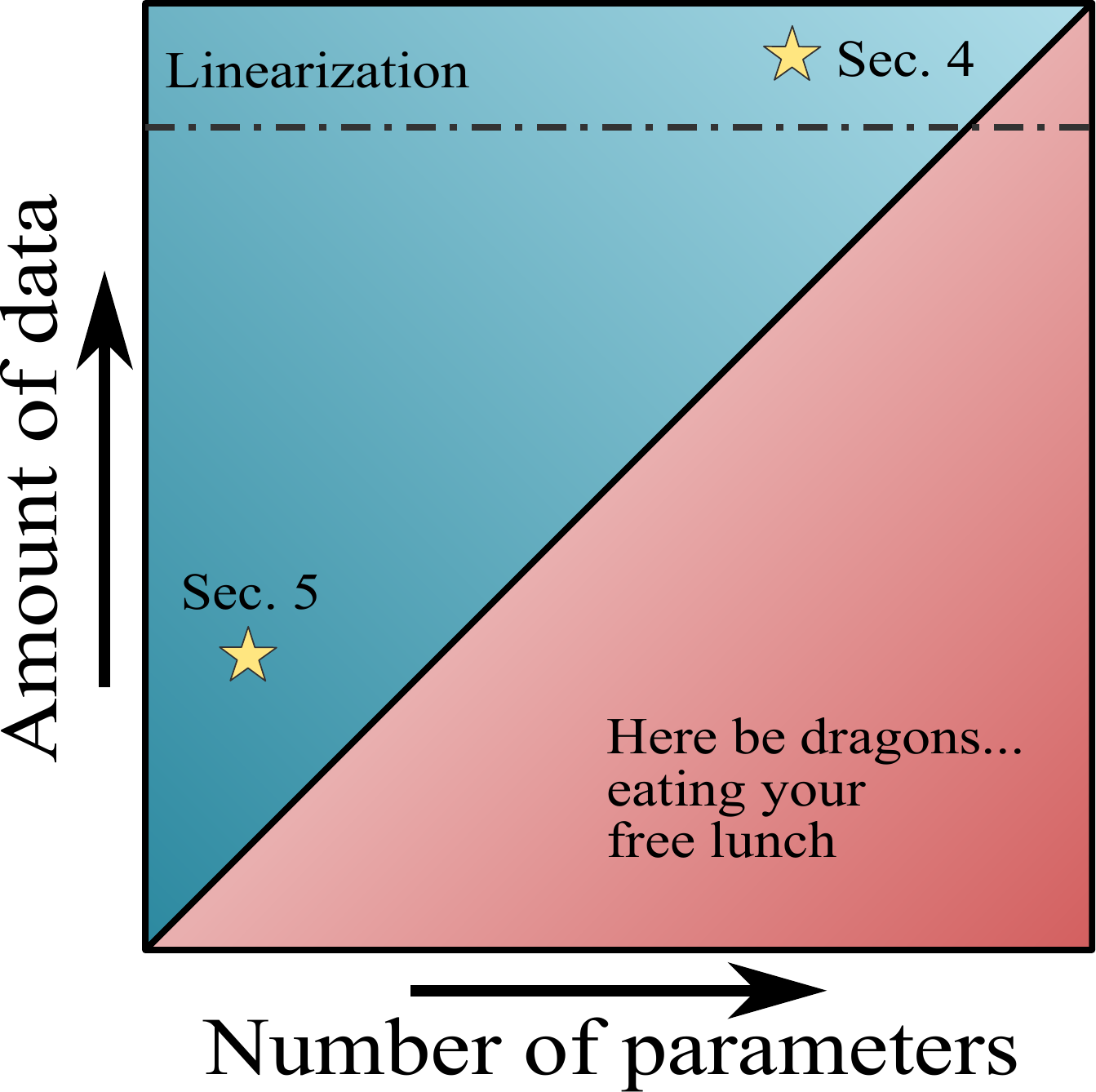}
	\end{center}
	\caption{\label{fig:cartoon}
		Estimation demands data. At some point (the exact location of which depends are far too many factors to quantify), the distribution of data becomes well-approximated by a Gaussian, allowing an effective linearization of the problem. This greatly simplifies the calculations required to solve the estimation problem. When this is not the case, the problem demands more resources and more clever numerical algorithms to approximate the solution. In any case, the more parameters one has in their model, the more data is required to learn anything.
	}
\end{figure}

We propose that many of these problems can be ameliorated, or at least properly quantified, using a statistically principled approach. Within the statistical phrasing of the problem we provide a Bayesian solution \cite{Bretthorst1988}, complete with a numerical implementation. We show the problem can be solved analytically in the limit of large of amounts of experimental data. At the other extreme---the small-data limit---a numerically stable Monte Carlo algorithm \cite{doucet_tutorial_2011} approximates the full Bayesian solution. Our two approaches provide a robust solution to the software side of the noise spectroscopy problem. These two regimes are schematically depicted in \autoref{fig:cartoon}. Though the physical model we consider is simplified to illustrate the novel aspects of this work, we note that our approach is very versatile. We discuss this later.

We summarize the performance of our approach for the large- and small-data limits in \autoref{sec:numerics}, with complete details including all source code in the supplementary material.
In the large-data limit, our approach can give almost an order of magnitude improvement in performance over state-of-the art estimation strategies.
By contrast, in the small-data regime, we can even achieve two orders of magnitude improvement with just 2,500 bits of data.
Thus, our approaches yield immediate and dramatic benefits in terms of experimental costs.

An expert reader may have realized that our work seems related to recent work of Zwick, Alvarez and Kurizki (ZAK) \cite{Kurizki_maximizing_2016}. It is useful to highlight the main differences between that work and ours. There the authors discuss the problem of designing experiments to maximize the expected information of select spectral properties (``bath parameters''). The main difference is that ZAK treat the single parameter case, and are thus able to obtain clever analytic solution to the experiment design problem. In contrast, our approach is inherently multidimensional and thus more general, but this means we only provide numerical algorithms. These works can thus be seen as complementary\footnote{Moreover, though ZAK suggest a Bayesian approach to the problem, they do not detail what numerical Bayesian updating algorithm is used. Our software is capable of realizing the protocol of ZAK for their specific problem and its multidimensional generalization.}. 

Our paper is organized as follows. In \autoref{sec:motivation} we motivate the problem from a physical perspective. In \autoref{sec:math} we extract the core mathematics, simplifying as much as possible the physical equations in order to phrase the problem as one of statistical estimation theory. We provide two solutions (as discussed in \autoref{fig:cartoon}) at varying degrees of complexity and economy. In \autoref{sec:gaussian_proceses} we provide the Gaussian process model which is valid in the large data limit. In \autoref{sec:hyperparameterized} the full Bayesian solution is outlined and applied to the ubiquitous $1/f$ noise model. A summary of the findings from our numerical experiments is given in \autoref{sec:numerics}. We wrap up in \autoref{sec:discussion} with a brief discussion. Full implementation details and code to reproduce the results are listed in the ancillary files for this paper. In particular, this paper can be seen as a \emph{living document} as follows: when the source is downloaded and run, new random variables will be drawn and the data will change. Therefore, the figures will also change, while---hopefully!---the conclusions drawn from them will not.

%=============================================================================
\section{Physical motivation}
\label{sec:motivation}
%=============================================================================

{In order to present our results, it is useful to introduce a concrete physical model that is simple enough that the Open Quantum Systems and Control language does not distract from the main features of our statistical approach, but that is sufficiently non-trivial to be a relevant model from the physical point of view.}

Our toy physical model is the dynamics of a two-level system (TLS) in the presence of dephasing noise generated by a zero-mean, Gaussian, stationary process. The TLS and its environment are assumed to be initialized in a factorizable state of the form $\rho_{SB} = \rho_S \otimes {\rho}_B$, with ${\rho}_B$ being the initial state of the bath. Furthermore, their dynamics is ruled by a Hamiltonian that, in the interaction frame with respect to the natural dynamics of the bath, takes the form 
\begin{align}
H(t) = \frac{\sigma_z}{2} \otimes B(t) + H_{\ctrl}(t),
\end{align}
where $B(t)$ represents the bath noise, and $H_{\ctrl}(t)$ is a control Hamiltonian acting solely on the TLS. We will describe the bath as if it is quantum but a classical bath is of course just a special case where all the bath operator that appear in our equations commute at different times, i.e. $[B(t_1),B(t_2)]=0$. For simplicity, we shall assume $H_{\ctrl}(t)$ to enact instantaneous $\sigma_x$ pulses at times $\{t_{i}\}$, i.e., $H_{\ctrl}(t) = \frac{\pi}{2} \sum \delta(t-t_i) \sigma_x$, such that, in the so-called toggling frame with respect to the control, the Hamiltonian takes the even simpler form $$ H(t) = y(t) \frac{\sigma_z}{2} \otimes B(t),$$
with $y(t)$ a binary function taking values in $\{-1,1\}$ and switching at times $\{t_i\}$. We will be interested in the expectation value of a Pauli operator $\sigma_\alpha$ at a time $T$, given by $\langle \sigma_\alpha(T)\rangle = {\rm Tr}[ U(T) \rho_{S} \otimes \rho_B U(T)^\dagger (\sigma_\alpha\otimes \mathbf{1}) ]$, with the unitary evolution given by the appropriate time ordered exponential $U(T) = \mathcal{T}_+ (e^{- i \int_0^T H(s) ds })$. It is important to highlight that when the bath $B(t)$ has a component $\beta(t)$ that is a classical stochastic process, one can only access the average expectation value of $\sigma_\alpha$ over many realizations of $\beta(t)$. That is one can measure $\langle \langle \sigma_\alpha(T) \rangle \rangle_c$ with $\langle \cdot \rangle_c$ denoting average over realizations of $\beta(t)$. To ease the notation we will denote simply by $\llangle \cdot \rrangle$ when both averages are taken. Under these conditions, the expectation value of an operator $\sigma_\alpha$ is given by{ ~\cite{paz-silva_multiqubit_2017}}
\begin{align}
\llangle \sigma_\alpha(T)\rrangle & =
		\Tr_S \!\! \left[
			\Tr_B \!\! \left[
				\left \langle
					\mathcal{T}_+ \Big( e^{ -i \int_{-T}^T \!\!\! dt \tilde{H}(t)} \! \Big)
				\! \right \rangle_c \!\! {\rho}_B
			\right] \rho_S \sigma_{\alpha}
		\right],
\end{align}
with $\tilde{H}(t) = \begin{cases} 
-\sigma_\alpha H(T-t) \sigma_\alpha &  \textrm{for} \,\,\,\,\,\,\,\,  0 \leq t \leq  T \\
\,\,\,\,\,\,\,\,\,\,\, H(T+t)  &  \textrm{for  } -T \leq t < 0 
\end{cases}$. The assumption that $B(t)$ is a Gaussian, zero-mean, stationary process, implies that only the second cumulant of the process is non-vanishing, i.e., $C^{(2)} (B(t_1) B(t_2)) =\Tr [\langle B(t_1) B(t_2) \rangle_c  {\rho}_B] = \Tr [\langle B(t + t_1-t_2) B(t) \rangle_c  {\rho}_B] \,\,\, \forall \,\,\, t$. In turn this leads to the exact solution, in both time and frequency domains, being given by{ ~\cite{Breuer:book, paz-silva_multiqubit_2017}}
\begin{widetext}
	\begin{gather}
		\begin{aligned}
		\llangle \sigma_\alpha (T) \rrangle & = 
				e^{
					- \frac{1}{2} \left(\frac{1 - f_{z,\alpha}}{2}\right)^2
					\int_{0}^T d t_1 \int_{0}^t d t_1 \,y(t_1)\, y(t_2)\,  C^{(2)}(B(t_1) B(t_2))
				}
 \langle \sigma_{\alpha}(0) \rangle , \\
			\label{eq:decoh}
			& = e^{ - \frac{1}{2} \left(\frac{1 - f_{z,\alpha}}{2}\right)^2
				\int_{-\infty}^\infty  \frac{d \omega}{2 \pi} F(\omega,T) S(\omega)  } \langle \sigma_{\alpha}(0) \rangle .
		\end{aligned}
	\end{gather}
\end{widetext}
Here $f_{\alpha,z} = \sigma_\alpha \sigma_z \sigma_\alpha \sigma_z$ takes values in $\{-1,1\}$, $F(\omega,T) = |F^{(1)}(\omega,T)|^2$ is the filter function, and $S(\omega) = \int_{-\infty}^\infty  dt  C^{(2)} (B(t ) B(0)) e^{- i \omega t}$ is the power spectrum of the noise process. Additionally, $F^{(1)}(\omega,T)= \int_0^T  dt \,  y(t) \, e^{i \omega t } $ is the so-called order-one fundamental filter function which depends purely on the control~\cite{Paz2014, Kofman:04}. These equations capture the dephasing behaviour of a qubit in the presence of a classical---as in the various semiclassical approximations often made in NV centers~\cite{bar-gill_suppression_2012} or NMR~\cite{Alvarez2011}---noise, or quantum noise---as in the case of a bosonic bath in a thermal state~\cite{CaldeiraLeggett:81, Breuer:book}. 

In this language, suppressing the decoherence is akin to minimizing the value of the exponent on the right hand side of Eq.~\autoref{eq:decoh}. This can be achieved via the use of control: one can choose a routine whose filter has a small overlap with the power spectrum. Dynamical decoupling (DD) sequences, for example, achieve this by generating filters which vanish at $\omega=0$ and are flat around it. They are thus very effective against decoherence generated by ``slow'' noise, where $S(\omega)$ is mostly supported around $\omega=0$. For noise with considerable high frequency contributions, the DD sequence can be ineffective or can, in analogy to the anti-Zeno effect~\cite{antiZeno}, even generate the opposite effect, i.e., decoherence enhancement. On the other hand, if information about the power spectrum is available then optimal control techniques can be used to find a control routine that minimizes the overlap and thus the decoherence~\cite{Paz2014, Kofman:04}. This is clearly the ideal situation, but it begs the question if the necessary information can be obtained in an Open Quantum system scenario.

Complete knowledge of the Hamiltonian describing the decoherence process and of the initial state of the environment would grant us perfect knowledge of the power spectrum and, in such situations, optimal control methods could be used to minimize the decoherence of the TLS. However, even when Gaussianity is imposed a priori~\cite{Breuer:book}, such knowledge is rarely available. For example, the temperature of the thermal state of a bosonic environment or the dispersion relation for the bosonic modes is usually unknown. Fortunately, as seen from the equations above, the decoherence process only depends on $S(\omega)$, i.e., not on the actual form of $B(t)$ or even $\rho_B$ but on the bath correlations they induce. This quantity, while not directly measurable can be inferred from the measurable (average) response of the TLS to different control Hamiltonians. 

This is the working principle behind recently proposed \emph{noise spectroscopy} protocols. Schematically, one such protocol would work as follows. Imagine preparing a $+1$ eigenstate of $\sigma_x$ at time $t=0$, in such way that the expectation value of the observable $\sigma_x$ at the final time $t=T$, given $H_{\ctrl}(t)$, is as in Eq.~\eqref{eq:decoh}. Different choices of ${H_{\ctrl}(t)}$ result in different filters $F(\omega,T)$, and different experimentally accessible values of $\llangle \sigma_x (T)\rrangle\vert_{H_{\ctrl}(t)}$. In principle, it should be possible to choose a sufficiently large set of different control sequences in such way that the integral in the exponent can be deconvolved, and information about $S(\omega)$ can be inferred. Different approaches to this problem, under different simplifying assumptions, have been proposed and even experimentally implemented~\cite{Bylander2011, Alvarez2011, Muhonen2014, Wang2017, Norris2016}. More exotic protocols have been proposed to characterize more general noise processes, such as non-Gaussian noise~\cite{Norris2016} or noise affecting multiple qubits~\cite{paz-silva_multiqubit_2017, Cywinski2}. While we will not consider them here in detail, we note that the statistically motivated methods can in principle be also used there via an appropriate generalization.

In the remainder, we will abstract as much physical detail as possible for brevity and generality. This allows us to easily apply techniques from statistical decision and estimation theory.

%=============================================================================
\section{Bayesian spectral estimation}
\label{sec:math}
%=============================================================================

In the physical description above, we make reference to observations as being the
average values of observables. By contrast, in real experiments, observations are made by acquiring single bits of data at a time through projective measurements
of single quantum systems. These two views of experimental observations agree only
in the limit that very large numbers of projective measurements are made on identical copies of the system.
Reasoning about noise spectroscopy in the presence of experimental
constraints is thus, at its core, a statistical problem not suited to the ``data-fitting'' paradigm we are more used to.
To make this precise we first extract the core mathematical elements of the problem.
Mathematically, we are interested in the exponent appearing in \autoref{eq:decoh},
\begin{equation}
	\label{eq:inner-product}
	\chi(S;F_j) = \frac1{2\pi} \int_0^\Omega S(\omega)F_j(\omega) d\omega,
\end{equation}
where $\Omega$ is a high frequency cut-off which is often imposed experimentally and required for the numerical integration we use. So as to not introduce too many complications, we assume that $\Omega$ is known. 

Recall that we do not have direct access to $\chi$ as it is only exposed experimentally through the statistical model in \autoref{eq:decoh}. Moreover, expectation values of observables also cannot be measured directly and will always come with fluctuations due to finite sample sizes. Thus, we prefer to work from the bottom up, considering the precise distribution of each bit of data. To this end, let $r$ be a the binary random variable with distribution
\begin{equation}
	\label{eq:model}
	\Pr(r = 1|S; F_j) = \frac{1}{2}\left(1+e^{-\chi(S;F_j)}\right),
\end{equation}
such that the expectation value in \autoref{eq:decoh} obeys
\begin{align*}
	\llangle \sigma_x \rrangle =  \Pr(r = 1|S; F_j) - \Pr(r = -1|S; F_j).
\end{align*}
This is the most fundamental statistical model and we should process data at this level whenever possible. But wait, what does it mean to \emph{process data}? This is where the Reverend Bayes comes in.

The notation $\Pr(A|B)$ is read ``the probability of $A$ being true given $B$ is known to be true''. So, $\Pr(r=1|S;F_j)$ is the probability of observing $r=1$ given the filter $F_j$ is used and the spectrum is $S$. Ah, but that seems a bit awkward, doesn't it? Isn't the spectrum the thing we don't know? To rectify this, we \emph{invert} the probability using Bayes' rule:
\begin{equation}
	\label{eq:bayes_rule}
	\Pr(S|r;F_j) = \frac{\Pr(r|S;F_j)\Pr(S|F_j)}{\Pr(r|F_j)}.
\end{equation}
Some terminology \cite{sivia2006data}: $\Pr(r|S; F_j)$ is called the \emph{likelihood function} and in physics it is always given by the physical model; $\Pr(r|F_j)$ is called the \emph{evidence} and it is usually ignored as it can be determined by normalization; $\Pr(S|F_j)$ is called the \emph{prior} and encodes the information we have about the spectrum before the data is take; and finally, $\Pr(S|r;F_j)$ is called the \emph{posterior}, which is the information we have about the spectrum \emph{after} the experiment---exactly what we want to know!

In general, performing this inversion is both analytically and computationally intractable. There are two general approaches to solving this problem. Either we make analytical approximations or we employ clever numerical integration techniques. Here we demonstrate both. But, the problem and solutions are also not decoupled from how much can be assumed known about the spectrum---the \emph{dimension} of model---and the amount of data available, such that the domain of applicability of each solution is restricted in subtle ways. This is shown pictorially in \autoref{fig:cartoon}.

%-----------------------------------------------------------------------------
\subsection{Big data: analytical approximations with weak assumptions}
%-----------------------------------------------------------------------------

In the large data\footnote{The definition of ``large'' is intentionally left ambiguous as it depends on far too many things to give a precise number to.} limit, we can appeal to the central limit theorem. In the Gaussian limit of the likelihood function we effectively linearize the model. For brevity, we will denote $\chi(S;F_j)\mathrel{=:}\chi_j$. Suppose the the number of binary samples taken per filter function used is $N$. Denote each binary sample $r_{ji}$ and 
\begin{equation}
	\hat y_j = \frac1N \sum_{i=1}^N r_{ji}.
\end{equation}
Then $\hat y_j$ is a binomial random variable with mean and variance given by
\begin{subequations}
\begin{align}
	\mathbb E[\hat y_j] &= \frac12\left(1 +e^{-\chi_j}\right) \text{ and} \\
	\mathbb V[\hat y_j] &= \frac1{4N}\left(1-e^{-2\chi_j}\right).
\end{align}
\end{subequations}
Consider the random variable $\hat \chi_j = - \log\left( 2\hat y_j-1\right)$. Taylor expanding about the mean---that is, about the variable $\hat y_j - \mathbb E[\hat y_j]$---we have
\begin{subequations}
\begin{align}
	\mathbb E[\hat \chi_j] &\approx \chi_j, \\
	\mathbb V[\hat \chi_j] &\approx \frac{e^{2\chi_j}-1}{N}.
\end{align}
\end{subequations}
Another way to specify data when this approximation is valid is to treat $\hat\chi_j\sim \mathcal N (\chi_j,\sigma_j^2)$, for each filter $F_j$, where 
\begin{equation}
 	\sigma_j^2 = \frac{e^{2\chi_j}-1}{N}.
\end{equation} 

Though this approximation will be valid when strong assumptions are made to reduce the model dimension on $S$, it's real utility is in allowing a tractable solution for weak assumptions on $S$. In \autoref{sec:gaussian_proceses}, we will specify precisely how we model $S$ when this approximation holds.

%-----------------------------------------------------------------------------
\subsection{Small data: sequential Monte Carlo}
%-----------------------------------------------------------------------------

For smaller data sets, the normality assumptions made in the previous section are difficult to justify, and may fail altogether.
Thus, we must take an alternative approach to calculating the posterior distribution given by \autoref{eq:bayes_rule}.
To do so, we note that the main advantage of the Gaussian approach of the previous section was that it will allow us to represent the prior and the posterior distributions as being different members of the same family of distributions.
A promising alternative approach, then, is to consider more general families of distributions, perhaps at cost of greater computational effort.
In particular, we will use the sequential Monte Carlo (SMC) approximation, which represents the distributions $\Pr(S | F_j)$ and $\Pr(S | r; F_j)$ appearing in \autoref{eq:bayes_rule} by weighted sums of $\delta$-distributions.
This approximation is very general, and will allow us to be much more general in our treatment of $\Pr(r | S; F_j)$.
In particular, using SMC will allow us to easily express models for spectral density functions that can be described using a small number of parameters, such as $1 / \omega^\alpha$ for an unknown power $\alpha$.

On the other hand, using sequential Monte Carlo forgoes the benefits of the analytic approximations described in the previous section, such that there is a natural tradeoff between the two approaches with the amount of data being taken, and with the form of the models under consideration.
We detail the sequential Monte Carlo--based approach and compare it to the Gaussian process approach in \autoref{sec:hyperparameterized}.

%-----------------------------------------------------------------------------
\subsection{Filter functions and ``naive'' estimator}
%-----------------------------------------------------------------------------

\begin{figure*}
	\begin{center}
		\includegraphics[width=0.98\textwidth]{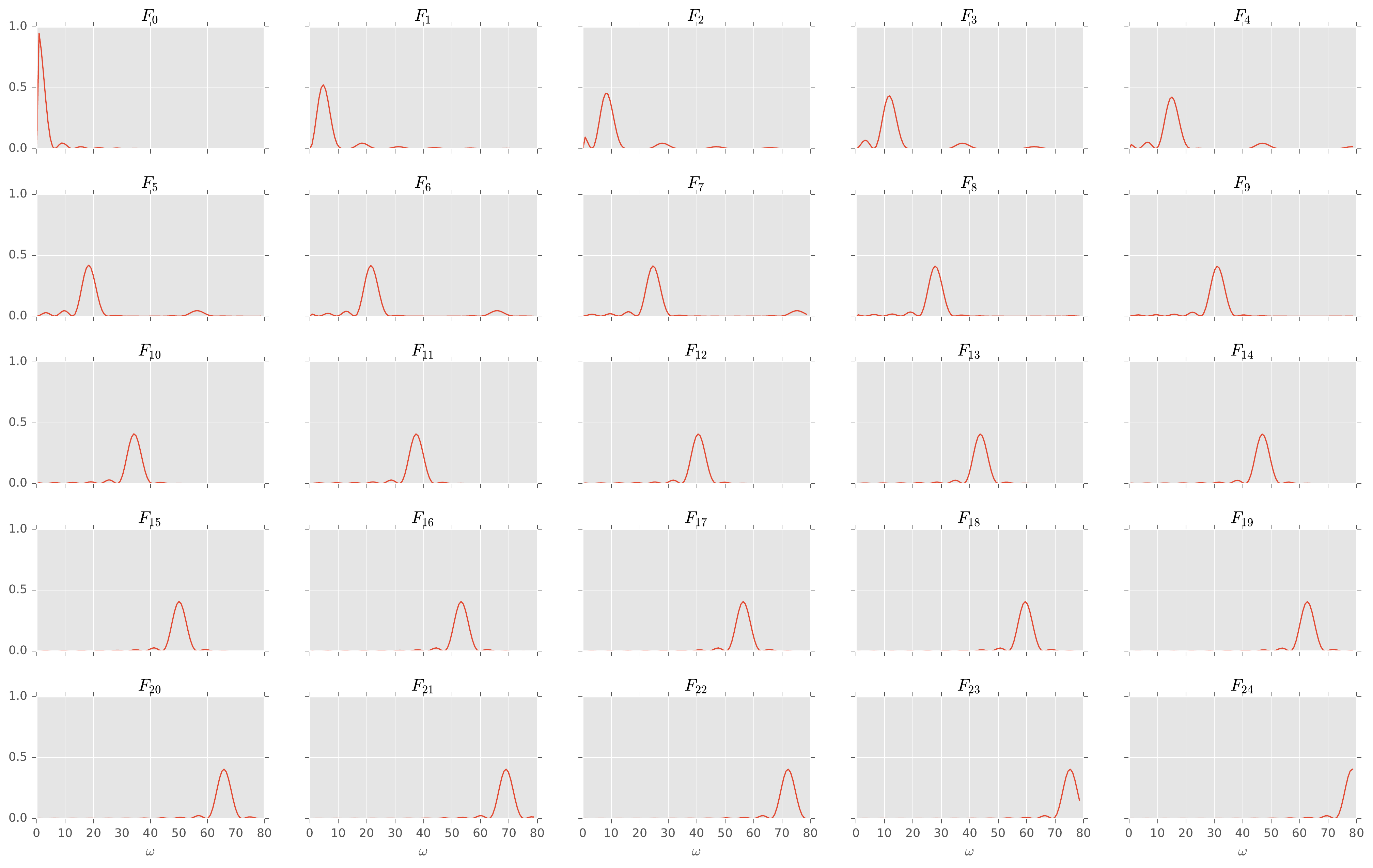}
	\end{center}
	\caption{\label{fig:filter-fns}
		The filter functions of the 25 sequences we consider in this work. 
	}
\end{figure*}

{For our numerical experiments we consider filters that arise from compressing so-called CPMG sequences~\cite{meiboom_modified_1958} of increasing number of pulses in a fixed total time $T$, as was done in Ref. \cite{alvarez_measuring_2011} for example.
For a CPMG sequence of $p$ pulses, $y(t)$ is a function which switches between $-1$ and $1$ at every pulse time $t_i =  (2 i+1) T/(2n)$. More involved sequence choices can be made, if one is interested in exploring higher frequency regimes, for example~\cite{norris_qubit_2016,paz-silva_multiqubit_2017}, but this simple choice is enough for our purposes. A useful feature of this choice is that it provides an intuitive way of producing filters whose main support is in a given frequency range. More specifically, the larger $p$ is the higher in frequency the main peak of $F(\omega)$ is. This is important becuase if no filter had support in a given frequency range it would be impossible to accurately estimate the power spectra in that regime.} We plot the filter functions used in this paper in \autoref{fig:filter-fns}.

\begin{figure}
	\begin{center}
		\includegraphics[width=0.98\columnwidth]{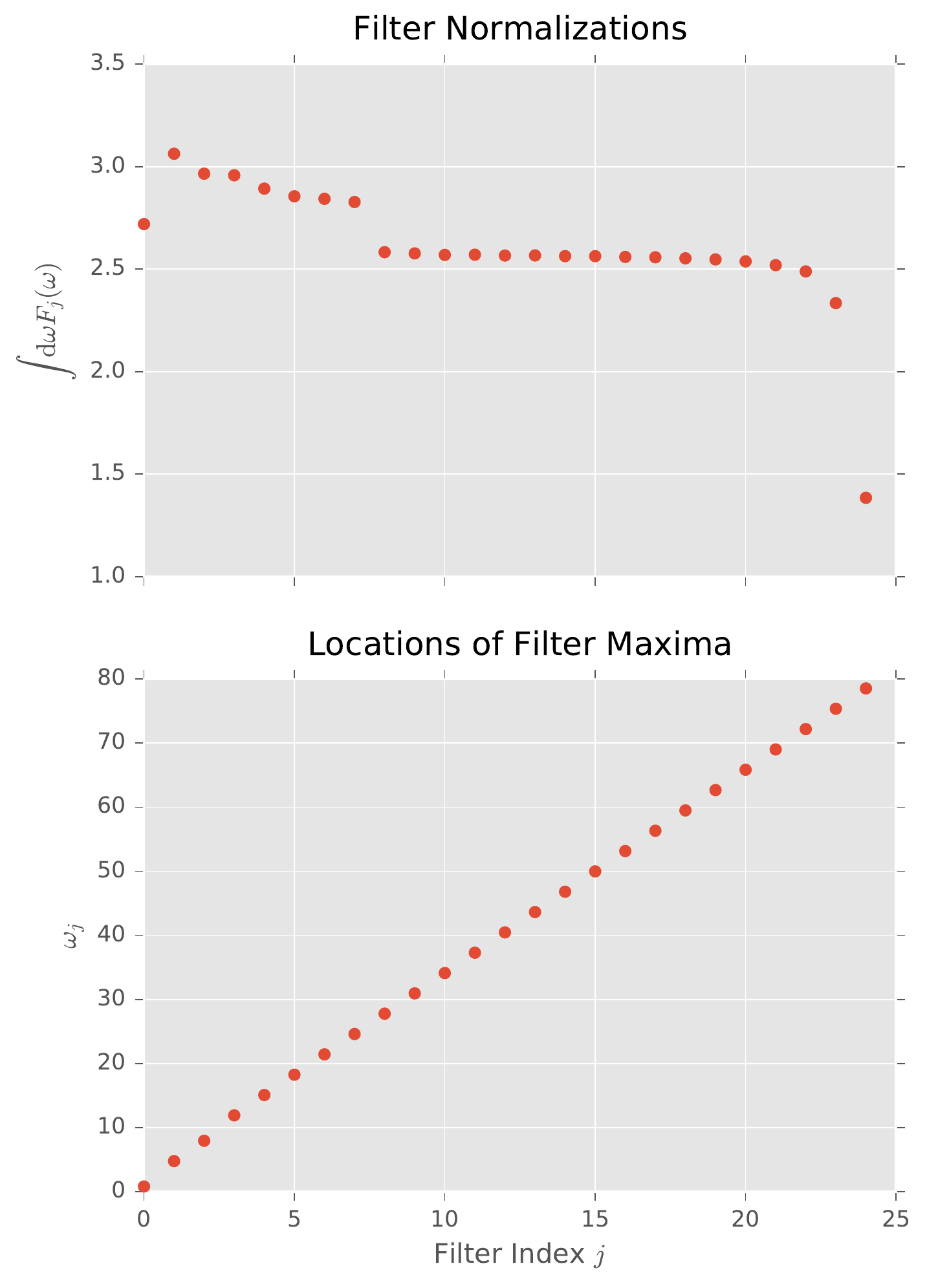}
	\end{center}
	\caption{\label{fig:filter-naive-props}
		Normalization $f_k$ (top) and peak frequency $\overline{\omega_k}$ (bottom)
		of each of the 25 filter functions in \autoref{fig:filter-fns}.
	}
\end{figure}

\begin{figure}
	\begin{center}
		\includegraphics[width=0.95\columnwidth]{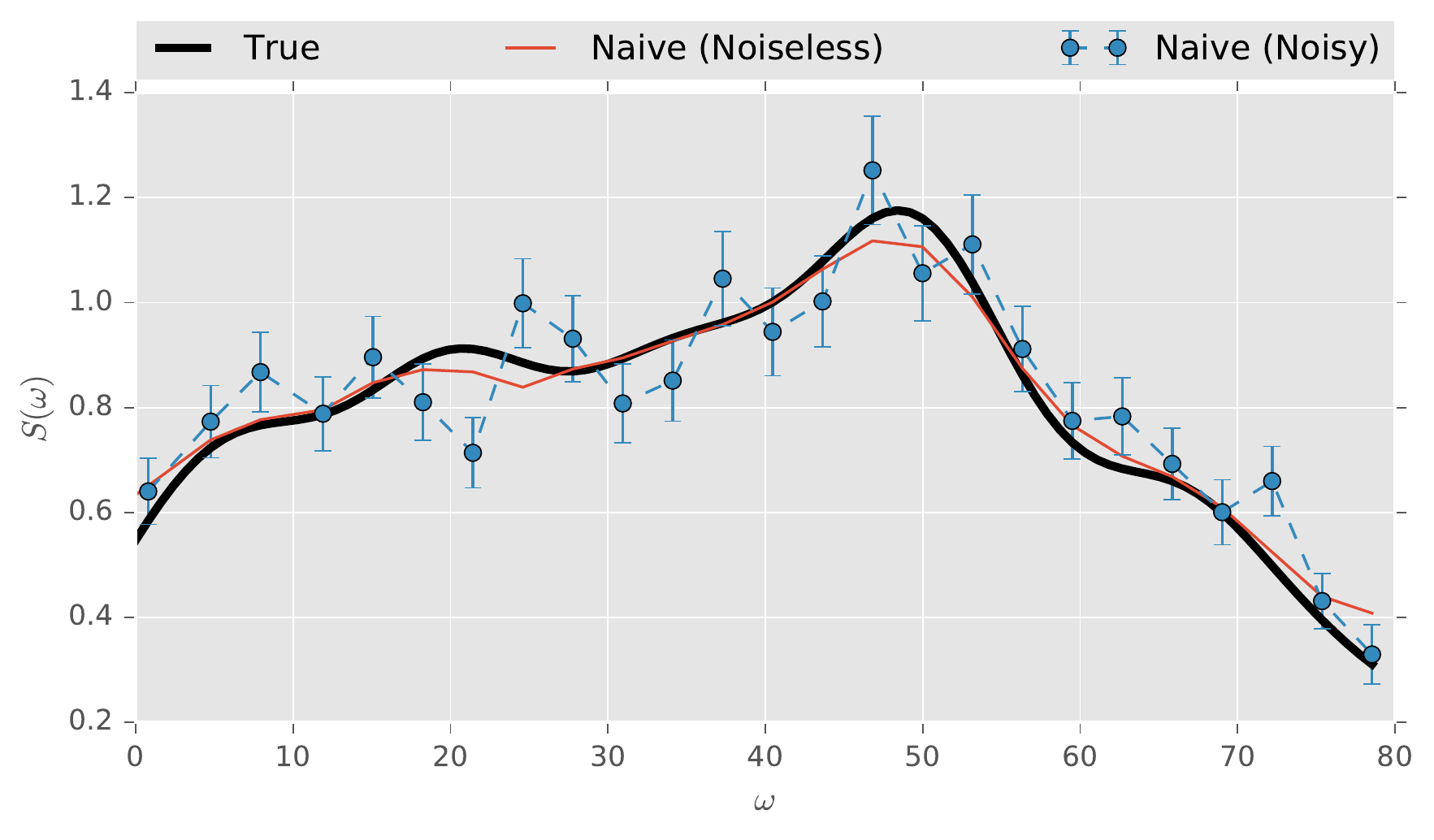}
	\end{center}
	\caption{\label{fig:naive-est}
		A randomly chosen true spectrum, compared to naive estimates
		from noiseless and noisy data. The data for each estimate is simulated
		using the 25 control sequences discussed in the main text,
		with the noisy data being simulated for $N = 1000$ repetitions per
		control sequence, and with the noiseless data being simulated for
		the limit of infinite repetitions per sequence. The discrepancy between the noiseless data curve and the true curve is due to the approximation of the filters as delta functions in the theory.
	}
\end{figure}

For the large data limit, we can use a simple data fitting estimator for a point of reference. 
In forming our naive spectral density estimate, we will require the normalizations $f_j \mathrel{:=} \int_0^\Omega \mathrm{d}\omega F_j(\omega)$ and the maxima $\overline\omega_j \mathrel{:=} \operatorname*{arg\,max}_{\omega} F_j(\omega)$. For the filters we consider (see \autoref{fig:filter-fns}), the normalizations and maxima are plotted in \autoref{fig:filter-naive-props}. Next, we approximate each filter function as
\begin{equation}
	F_j (\omega) \approx \delta(\omega - \overline\omega_j) f_j.
\end{equation}
This leads to
\begin{equation}\label{eq:chi}
	\chi_j \approx \frac{S(\overline\omega_j) f_j}{2\pi}.
\end{equation}
Suppose we identify the experimentally observed random variables $\hat \chi_j$ with the theoretical values $\chi_j$. Then, inverting \autoref{eq:chi}, we define the naive estimate 
\begin{equation}\label{eq:naive-est}
	\hat{S}_{\text{naive}}(\overline\omega_j) = \frac{2 \pi \hat{\chi}_j}{ f_j}.
\end{equation}
This will be compared to some more sophisticated, but more computationally expensive estimators. An example output of the naive estimator is shown in \autoref{fig:naive-est}.

%=============================================================================
\section{Big data: Gaussian process regression}
\label{sec:gaussian_proceses}
%=============================================================================

Suppose we are in the large data limit. That is, $N$ is big enough that all distributions are roughly Gaussian. To deal with the notion of a prior, or measure, on functions we treat the unknown spectrum as a random function $S(\omega)$. Denote the distribution of $S$ as $\Pr(S)$. To specify this concretely we discretize the support of the distribution to the set $W= \{\omega_1,\ldots,\omega_M\}$. This means that, whenever a numerical calculation is performed, we really only consider random variables $S_k \defeq S(\omega_k)$, which can be represented collectively as the vector $\vec{S}$.

\begin{figure}
	\begin{center}
		\includegraphics[width=0.95\columnwidth]{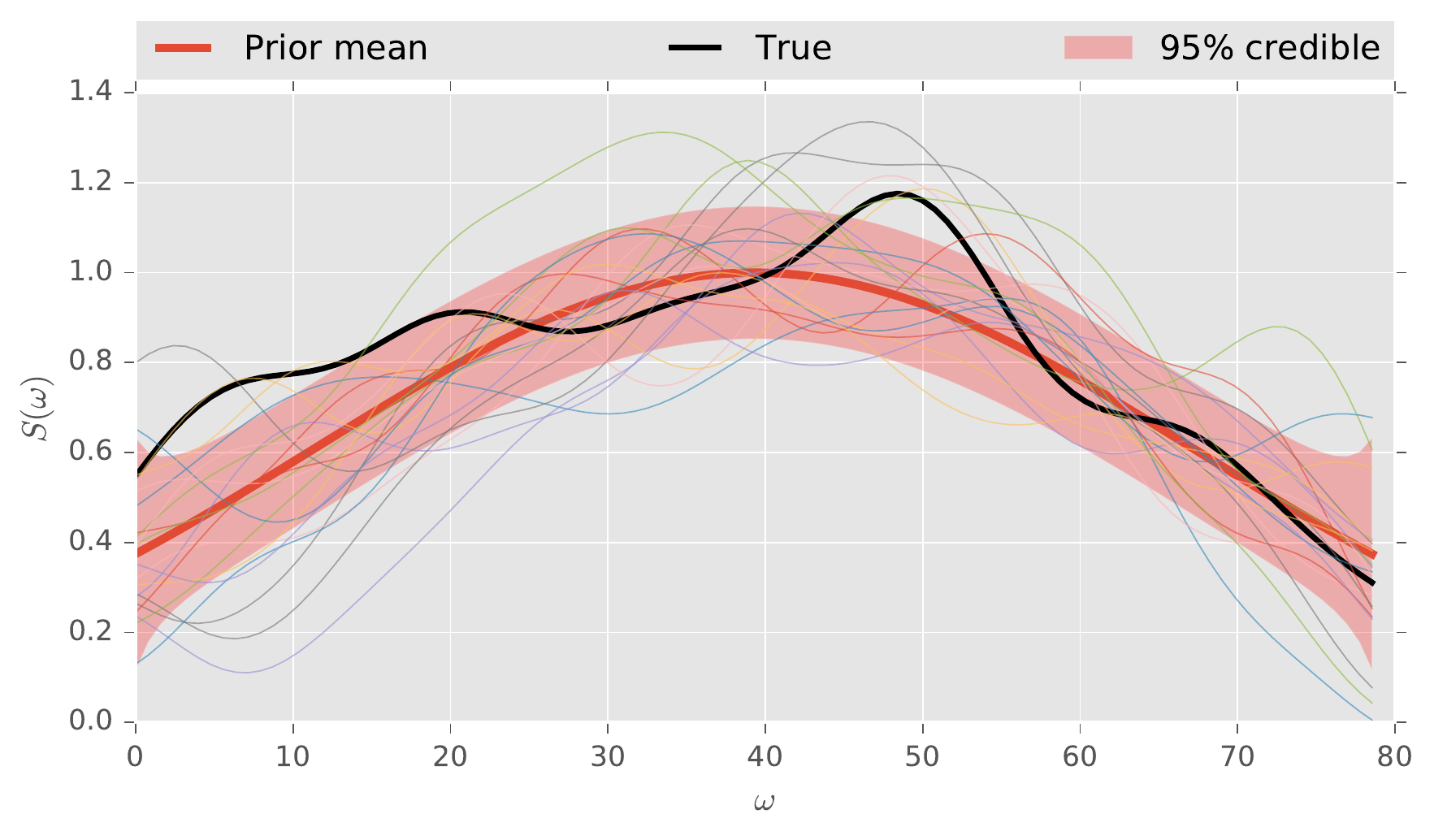}
	\end{center}
	\caption{\label{fig:example-prior}
		A visualization of a Gaussian Process. Here the mean function $\mu$ is taken to be a Gaussian function and we use the squared exponential kernel in \autoref{eq:sq-exp-kernel} with parameters $\kappa = 0.02$ and $\delta = 100$. In red, the mean and 95\% credible band is plotted. The other curves are samples from this GP. One of them, in solid black, we take to be the true spectrum. 
	}
\end{figure}

The simplest non-trivial distribution is Gaussian (or normal): $\vec{S} \sim \mathcal N(\vec{\mu}, \vec{k})$, where $\vec{k}_{jk}:= k(\omega_j,\omega_k)$ and $\vec{\mu}_j:= \mu(\omega_j)$ are the covariance and mean.
In the function space picture we write this $S(\omega)\sim \GP(\mu(\omega),k(\omega,\omega'))$, where $\GP$ stands for \emph{Gaussian process} \cite{rasmussen_2005_gaussian}, $\mu$ is the \emph{mean function} and $k$ is the covariance function, or \emph{kernel}. In standard notation,
\begin{align}
	\mu(\omega) &= \expect_S[S(\omega)] \text{ and} \\
	k(\omega,\omega') &= \expect_{S}[(S(\omega)-\mu(\omega))(S(\omega')-\mu(\omega'))].
\end{align}
In principle we can choose any functions $\mu$ and $k$ as our mean and kernel functions. However, there are natural choices and ones that have been found to perform well in a broad range of problems. The most common kernel is the so-called squared exponential\footnote{I know, like we need yet another Gaussian function. In fact, here is a little quiz: a Gaussian function describe five different things in paper. Only 1 in 10 MIT graduates can name them all. How many can you get?},
\begin{equation}\label{eq:sq-exp-kernel}
k(\omega,\omega') = \kappa e^{-\frac{(\omega-\omega')^2}{\delta}},
\end{equation}
where $\delta$ is a hyper-parameter which controls the correlation in $S$ for nearby $\omega$ and $\kappa$ controls the overall prior uncertainty. In a purely Bayesian context, we should have \emph{a priori} values for $\mu$, $\kappa$, and $\delta$. In other words, we believe the ``true'' spectrum is drawn according to a GP with these parameters. If this is the case, then no more needs to be done. If not, we would need to perform \emph{model selection} \cite{edwards_bayesian_1963}.
The topic of model selection is beyond the scope of this work and so we will chose specific values for $\mu$, $\kappa$, and $\delta$. To get some intuition for how this relates to qubit noise spectra, we have plotted a visualization of the GP prior we will use in \autoref{fig:example-prior}.

If we begin with a GP prior and the distribution of data is also Gaussian, then the posterior is Gaussian and we can derive an analytic expression for its mean function and kernel. As discussed above, in the large data limit we can do just this. Moreover, since each experiment is uncorrelated, we can process the data at once treating $\hat{\vec{\chi}}\sim \mathcal N(\vec{\chi}, \matr{\Sigma} )$, where $\matr{\Sigma}$ is a diagonal covariance matrix with entries given by $\sigma_j^2$.

Since the prior is Gaussian and the likelihood function is Gaussian, the posterior is also Gaussian. Determining its mean and covariance is a simple exercise in multivariate completing the square. Denote $\matr{G}$ as the matrix with entries $G_{k j} = F_j(\omega_k)(\omega_k-\omega_{k-1})/4\pi$, such that the trapezoidal rule applied to $\chi_j$ is written 
\begin{align}
	\chi_j &\approx \frac 1{4\pi} \sum_{k} F_j(\omega_k)S(\omega_k)(\omega_k-\omega_{k-1}),\\
	&=\sum_{k} {G}_{jk}S_k.\label{eq:chi_j-trapz}
\end{align} 
Then, Bayesian updating amounts to updating the covariance and mean as follows \cite{rasmussen_2005_gaussian}:
\begin{align}
	\vec{k} &\mapsto \vec{k}' = \matr{G}^{\rm T} \matr{\Sigma}^{-1} \matr{G} +\vec{k}^{-1} ,\label{eq:gp-cov-update}\\ 
	\vec{\mu} &\mapsto \vec{k}'^{-1}\left(\vec{\chi}^\T \matr{\Sigma}^{-1} \matr{G} +\vec{\mu}^{\T}\vec{k}^{-1} \right).\label{eq:gp-mean-update} 
\end{align}

\begin{figure}
	\begin{center}
		\includegraphics[width=0.95\columnwidth]{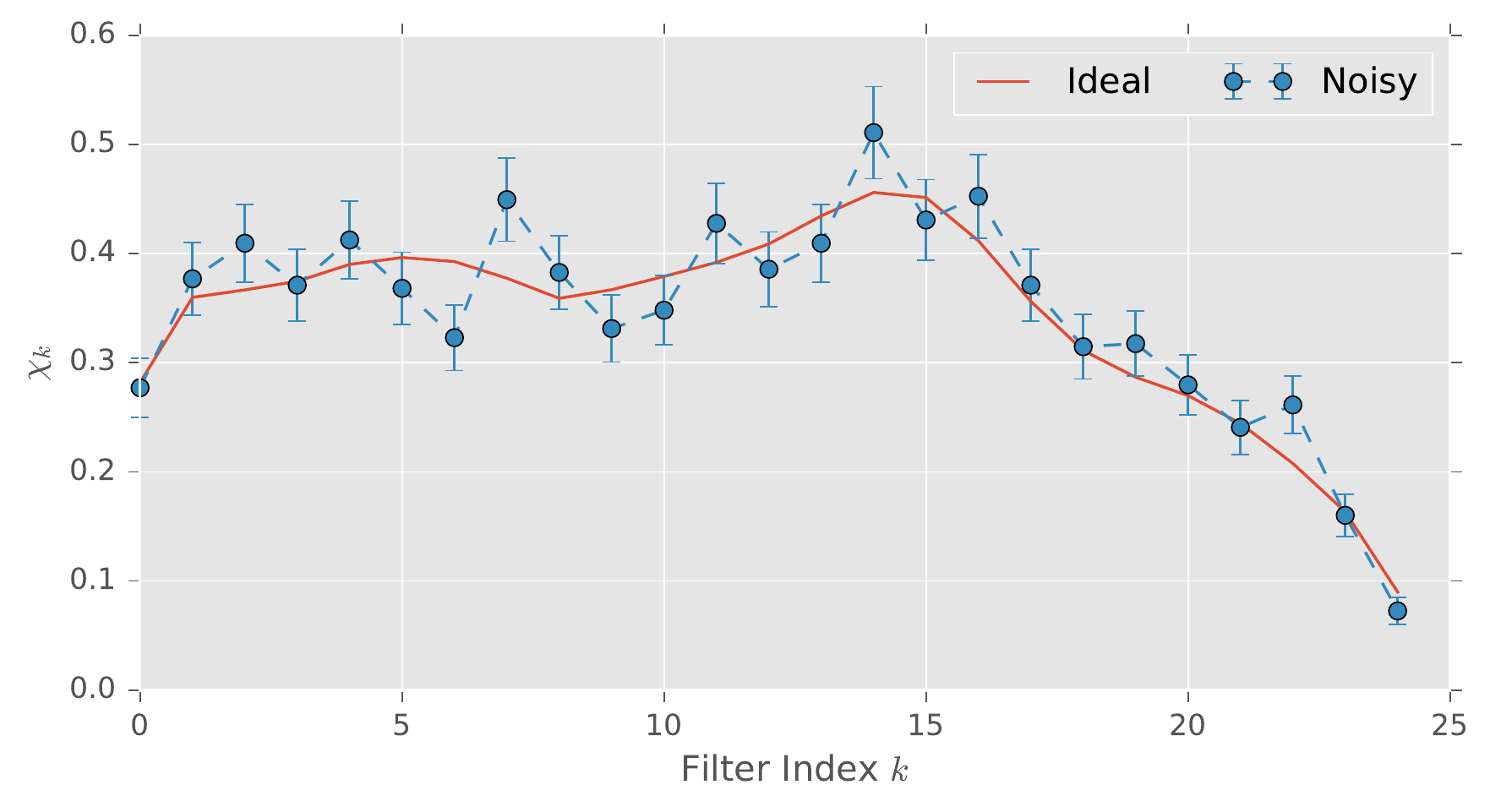}
	\end{center}
	\caption{\label{fig:example-naive-data}
		For $N = 1000$ repetitions, and using the true spectrum in \autoref{fig:example-prior}, we plot the observed data superposed over the theoretical (infinite precision) $\chi$'s. 
	}
\end{figure}

Let's take a look at an example simulation and use this GP estimator to find the spectrum. First, in \autoref{fig:example-naive-data} we plot the simulated data for $N = 1000$ repetitions per experiment.

\begin{figure}
	\begin{center}		
		\includegraphics[width=0.95\columnwidth]{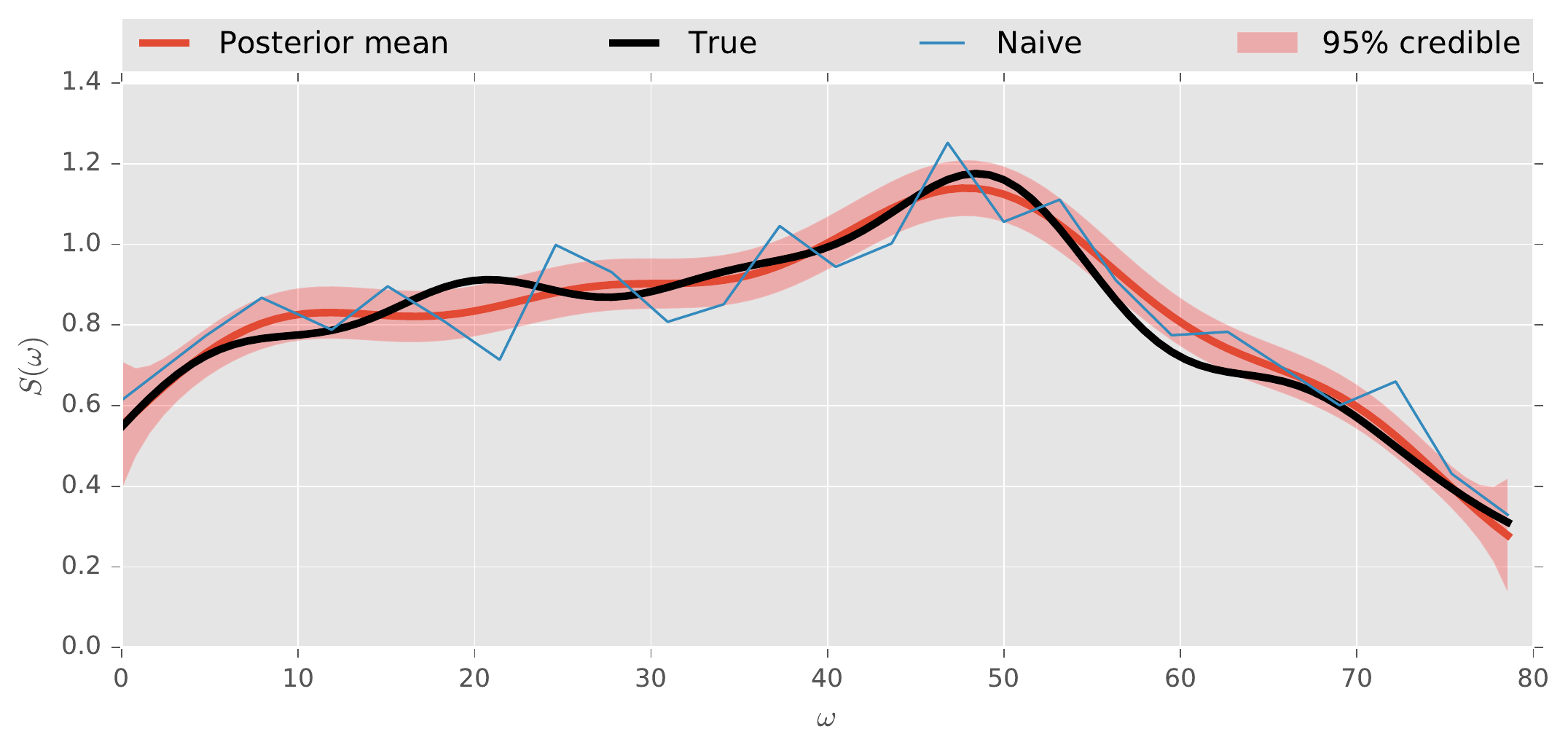}

		\includegraphics[width=0.95\columnwidth]{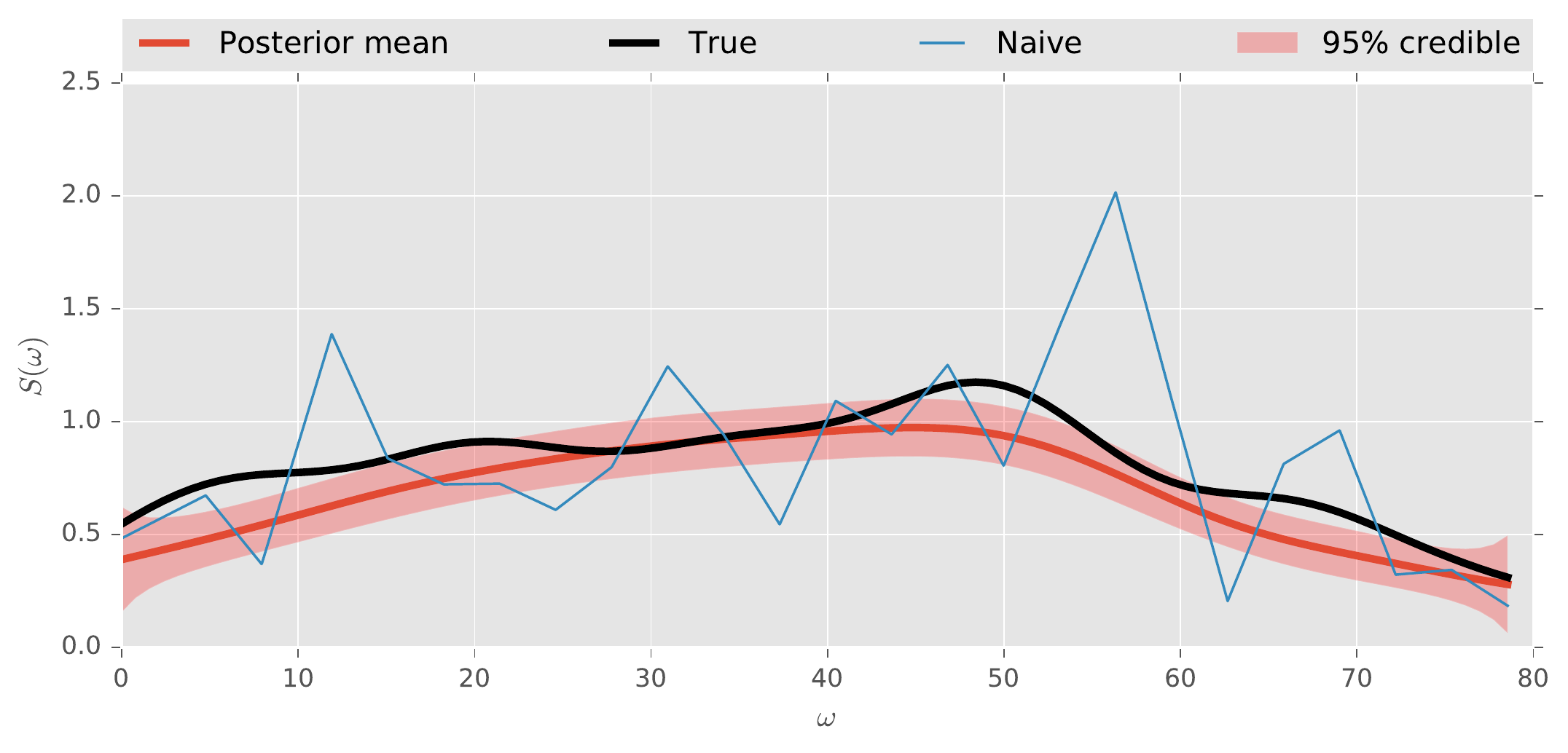}
	\end{center}
	\caption{\label{fig:example-gp-est}
		The posterior Gaussian Process, naive estimator and true spectrum using the data plotted in \autoref{fig:example-naive-data} (top), and with similar data taken using 100 shots per filter function (bottom).
	}
\end{figure}

Using the data plotted in \autoref{fig:example-naive-data}, we apply Equations \eqref{eq:gp-cov-update} and \eqref{eq:gp-mean-update} to get the posterior Gaussian Process. We plot the mean function and 95\% credible band in \autoref{fig:example-gp-est}. We see that the naive and GP estimator agree reasonably well when plenty of data is available. We also plot the same procedures for much less data ($N = 100$), where it is evident that naive estimator fails completely. Whereas, the GP estimator correctly hedges its bets by not suggesting any extreme features deviating the prior GP---it ``knows'' it doesn't have enough data to do so. A more extensive analysis of this difference is presented in \autoref{sec:numerics}.

\begin{figure}
	\begin{center}
		\includegraphics[width=0.95\columnwidth]{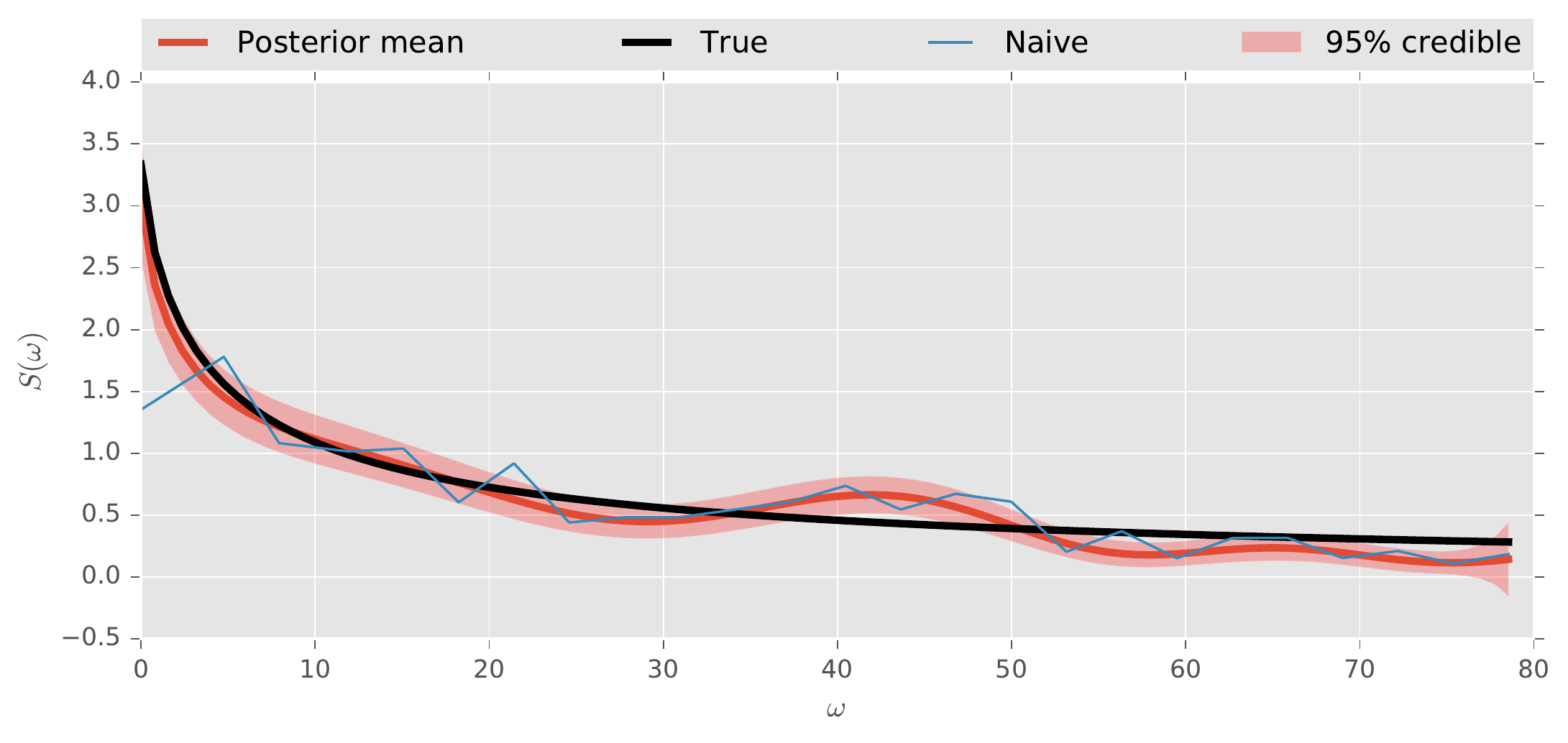}
	\end{center}
	\caption{\label{fig:example-gp-est-one-on-f}
		The posterior Gaussian Process and true spectrum for a $1/f^\alpha$ model
		\autoref{eq:one-on-f-model}, demonstrating that the Gaussian Process
		formalism does not consider the additional structure provided by the
		spectral model. For this simulation, each experiment was repeated $N = 50$ times
		and $\alpha$ was chosen uniformly at random in the interval $[1/2,1]$. ($A = 10, c = 3$.)
	}
\end{figure}

The GP estimator is not a silver bullet, however. First, as already noted, it is only valid when the data are drawn according to a Gaussian distribution---for example, when the Central Limit Theorem applies. Second, unless augmented with more sophisticated machine learning algorithms, the GP estimator can only reliably estimate features with sizes the order of $\delta$. To illustrate this, 
consider now the one-on-$f$ model (or, more generally the $1/f^\alpha$ model, with $f$ and $\omega$ being interchangeable here). This corresponds to a spectrum,
\begin{align}
	\label{eq:one-on-f-model}
	S(\omega; A, \alpha, c) =\frac{A}{\omega^\alpha + c},
\end{align}
where $c$ gives an effective low frequency cut-off. In this case, the spectrum has a high amount of structure. Indeed, the entire functional form is dictated by only a few parameters. We don't expect, then, that with the freedom allowed by the GP, it will be able to find this structure without a large amount of data. In \autoref{fig:example-gp-est-one-on-f}, we see that the GP is blind to the structure in the $1/f^\alpha$ model and estimates additional non-existent features. While the naive estimator also suffers from this problem, in the next Section we will show how to incorporate information about this  structure by the method of hyperparameters. This provides an alternative approach for cases in which a global property of the spectrum is
of more experimental interest than the spectrum itself.

After we describe the final estimator, we will provide an extensive numerical comparison of all estimators in \autoref{sec:numerics}.

%=============================================================================
\section{Small data: Hyperparameterized nonlinear regression}
\label{sec:hyperparameterized}
%=============================================================================

In the $1 / f^\alpha$ example discussed in the previous Section, the prior
uncertainty
was concentrated on a small number of parameters, such that the distribution of the spectra
at each point is very highly correlated. That is, if we perfectly knew the parameters
$A$, $\alpha$, and $c$ as they appear in \autoref{eq:one-on-f-model},
we would be able to precisely predict the value of $S(\omega)$ at arbitrary
$\omega$.

Though we can describe these correlations using the
methods of the previous Section, such that the covariance kernel is
effectively a low-rank linear operator, it can also be very useful
to describe our learning problem more directly. For example, in
\autoref{eq:one-on-f-model}, we can interpret
$(A,\alpha,c)$ as a vector of parameters in its own right,
as this vector describes the distribution over the alternative parameterization
implied by our discretization of $S(\omega)$.

In making this interpretation, we will use the method of hyperparameters
to incorporate our knowledge of an appropriate functional form for spectra into
our estimation model directly. This is the standard approach in Hamiltonian parameter estimation, for example, where prior knowledge of the physics allows for a drastic reduction in model dimension \cite{granade_robust_2012}.
Bayesian inference can be then applied directly to
the such hyperparameters instead
of at the level of the bare physical model. 

Generally, if the likelihood
function $\Pr(r | \vec{\eta})$ depends on a model vector $\vec{\eta}$ that itself
is distributed as $\Pr(\vec{\eta} | \vec{\theta})$ for some other vector $\vec{\theta}$,
then we can consider the marginalized
distribution
\begin{subequations}
	\label{eq:hyperparam-marginalization}
	\begin{align}
		\Pr(r | \vec{\theta}) & =
			\expect_{\vec{\eta} | \vec{\theta}} \left[
				\Pr(r | \vec{\eta}, \vec{\theta}) 
			\right] \\
		& =
			\int_{\supp{\vec{\eta}}} \dd\vec{\eta} \Pr(r | \vec{\eta}) \Pr(\vec{\eta} | \vec{\theta})
	\end{align}
\end{subequations}
as a likelihood function in its own right.

Returning to the problem of spectral density estimation, we note that
we can readily define the model vector $\vec\eta$ by
the inner product $\langle S, F_k \rangle = \chi(S; F_k) = \frac{1}{2\pi}\int S(\omega) F_k(\omega) \dd\omega$,
as predicting each inner product $\chi_k$ is sufficient to reproduce the entire
likelihood function. From this perspective, if we can reproduce each $\chi_k$
from a lower-dimensional model $\vec{\theta}$ (that is, a model with less parameters
than the number of filter functions used to gather data), then $\vec{\theta}$ represents
a more efficient hyperparameterization than taking $\vec{\eta}$ directly.
For example, consider hyperparameterizing
\autoref{eq:inner-product} following a $1/f^\alpha$ model,
\begin{align}
	\label{eq:chi-1f}
	\chi_k(\vec\theta) = \int_{0}^{\Omega}
		\frac{A F_k(\omega)}{\omega^\alpha+c} \dd\omega,
\end{align}
for a given ultraviolet cutoff $\Omega$ and with $\vec{\theta} = (A, \alpha, c)$.
This model has been studied experimentally, in particular
as a diagnostic for superconducting and spin qubits
\cite{yan_rotating-frame_2013,dial_charge_2013},
such that improvements even in this simple example immediately
yield experimental benefits.

\begin{figure}
	\begin{center}
		\includegraphics[width=0.95\columnwidth]{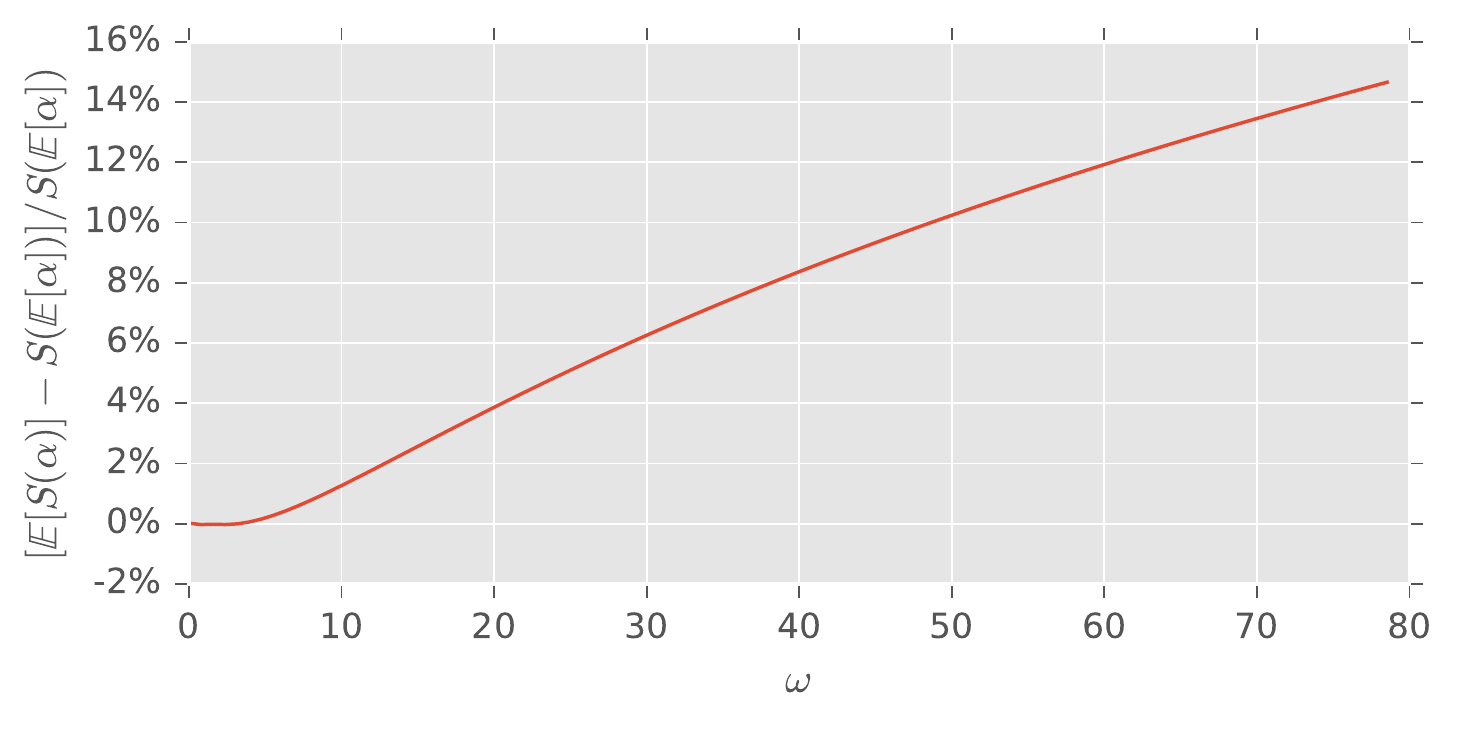}
	\end{center}
	\caption{\label{fig:mean-then-spectra}
		A $1/f^\alpha$ spectrum evaluated at the estimated $\hat{\alpha}
		= \expect[\alpha]$ compared with the Bayes estimate of the spectrum
		$\hat{S} = \expect[S(\alpha)]$. In this example, the hyperparameters are chosen according to \autoref{eq:one-on-f-cond}.
	}
\end{figure}

By expressing the estimation problem in terms of the hyperparameters
$\vec{\theta}$, we introduce a subtle distinction in how we report
our final estimates $\hat{S}$ once we have obtained a
datum $r$. We can report the spectrum
evaluated at the estimated hyperparameters $\hat{\vec{\theta}}
= \expect[\vec{\theta} | r]$ for our recorded data.
This estimate achives the best possible mean-squared error (MSE)
for reporting
the hyperparameters themselves, but does not necessarily
provide the best estimate of $S$.
As an alternative, we can instead report the Bayesian mean estimate
of the spectrum directly,
\begin{align}
	\hat{S}(\omega)
		& = \expect_{\vec{\theta}}[S(\omega; \vec{\theta}) | d].
\end{align}
That is, by taking the spectra and then the mean, we obtain the Bayesian
estimate of the spectrum, using our knowledge of the hyperparameters.
Though these two methods coincide for spectrum models that are linear
functions of their hyperparameters, for models such as $1/f^\alpha$,
the difference can be quite significant, as demonstrated in
\autoref{fig:mean-then-spectra}.

Critically, both methods for estimating $S$ from a posterior over
hyperparameters can be generated easily from the same data without additional
analysis. Thus, we are free to report the optimal estimate for questions of experimental
interest, rather than assuming \emph{a priori} that only one question will be asked
of our data.

In any case, we treat the spectrum more generally as being drawn from a parameterized distribution
of spectra, such that
$\Pr(S(\omega)|\vec{\theta}) = \delta(S(\omega)-S_{\vec{\theta}}(\omega))$
and $\vec{\theta}$ is a real-valued vector
for a functional form $S_{\vec{\theta}}(\omega)$ such as the
$1 / f^\alpha$ model discussed above. Therefore, once we have specified
$\vec{\theta}$, we have specified the unknown spectrum.
Following Bayes' rule \autoref{eq:bayes_rule}
as usual gives us a posterior distribution over the parameters $\vec{\theta}$,
conditioned on a data record $r$,
\begin{equation}
	\label{eq:hyper-bayes}
	\Pr(\vec{\theta}|r) = \frac{\Pr(r|\vec{\theta})\Pr(\vec{\theta})}{\Pr(r)}.
\end{equation}
Although the denominator looks like an innocuous normalization constant, producing accurate estimates of $\vec{\theta}$ requires its calculation.
Since we do not assume that $S_{\vec{\theta}}$ is a linear function of $\vec\theta$, and since we are concerned with efficiently utilizing small amounts of data, the analytic solution in terms of Gaussian process regression used for the process-model case cannot be directly applied here.
In lieu of that, our preferred method is sequential Monte Carlo \cite{doucet_tutorial_2011}, also known as particle filtering.
This algorithm computes posterior distributions of the form
given as \autoref{eq:hyper-bayes} by evaluating the likelihood at each of
many different \emph{particles}, each of which represents a particular
hypothesis about the true model vector $\vec{\theta}$ and an associated weight.
Expectation values over the posterior can then be replaced by finite sums over
the particles in the sequential Monte Carlo approximation.

We will use the implementation provided by the
QInfer package for Python \cite{granade_qinfer:_2016}.
In the following Section, we detail and present results obtained from a QInfer model for hyperparameterized spectral density estimation, and compare these results to those obtained from the Gaussian process regression method of \autoref{sec:gaussian_proceses}.
Our QInfer model will depend on the specification of a set of test frequencies $W = \{\omega_1, \dots, \omega_M\}$ and a spectral model function $S(\omega; \vec{\theta})$, where $M$ specifies the resolution of test frequencies.
The inner products $\langle S(\omega; \theta),
F_k(\omega) \rangle$ will then be approximated by numerically evaluating the
integral \autoref{eq:inner-product} in terms of the trapezoidal
rule \autoref{eq:chi_j-trapz} applied to the integrand $\{S(\omega_i; \theta) F_k(\omega_i) : \omega_i \in W\}$.
This design allows for our model to be very general with
respect to the particular choice of $S(\omega; \vec{\theta})$. 

In this paper, we will work with one such description by writing the $1/f^\alpha$ model as a \emph{hierarchal model} in which $r$ is a random variable that is defined by its distribution conditioned on our new hyperparameters $\vec{\theta}$.
In particular, the conditional distribution of $r$ is given by\footnote{An exponentially distributed random variable is denoted $x\sim {\rm Exponential}(x_l,\lambda)$ and has a probability density function $f(x) = \lambda \exp(-\lambda (x-x_l))$ for $x>x_l$.}
\begin{subequations}
	\def\theparentequation{\arabic{parentequation}}%
	\label{eq:one-on-f-cond}
	\begin{align}
		\tag{\ref*{eq:one-on-f-cond}}
		r|A, \alpha, c &\sim {\rm Bernoulli}\left( \frac{1+\e^{-\chi_k(A,\alpha,c)}}{2}\right),\\
		\intertext{
			with our prior on the hyperparameters $\vec{\theta} = (A, \alpha, c)$ given by
		}
		% This is a massive hack to let us align mixed equations and subequations that include intertext.
		\addtocounter{parentequation}{1}
		A & \sim {\rm Normal}(10,0.025),\\
		\alpha & \sim {\rm Uniform}([0.5,1]),\\
		\text{and } c & \sim {\rm Exponential}(0.1,3).
	\end{align}
\end{subequations}
For more details, please see the complete source code provided in
the supplemental material.

%=============================================================================
\section{Numerical experiments}
\label{sec:numerics}
%=============================================================================

We have already demonstrated some comparisons between the different approaches in the previous sections. The purpose of this section is to consolidate and expand on those illustrative comparisons. Namely, we will compare the performance of the naive, GP and hyperparameter estimator over many randomly chosen spectra. The comparison is facilitated by the mean squared error metric. Let $\hat S$ be an estimate of the true spectrum $S$. Then, the error---or \emph{loss}---is defined as
\begin{equation}
	L(S,\hat S)	= \int_0^\Omega |S(\omega) -\hat S(\omega)|^2 d\omega.
\end{equation}
To compare different estimators, we select a true spectrum randomly from the prior, simulate experiments, calculate the estimators, record the loss of each, and repeat. Then we plot a histogram of the achieved accuracy.

\begin{figure*}
	\begin{center}
		\includegraphics[width=0.48\textwidth]{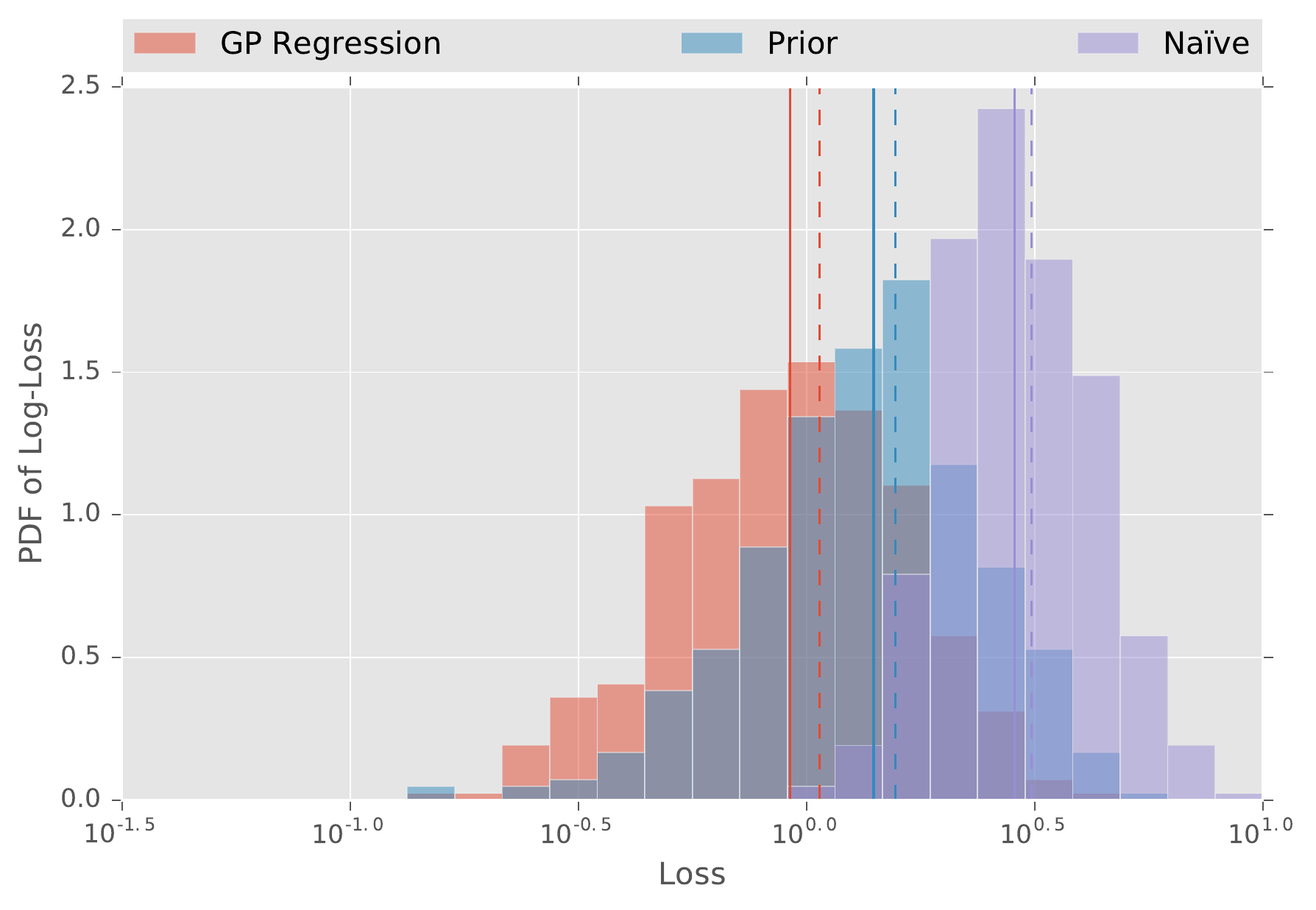}
		\includegraphics[width=0.48\textwidth]{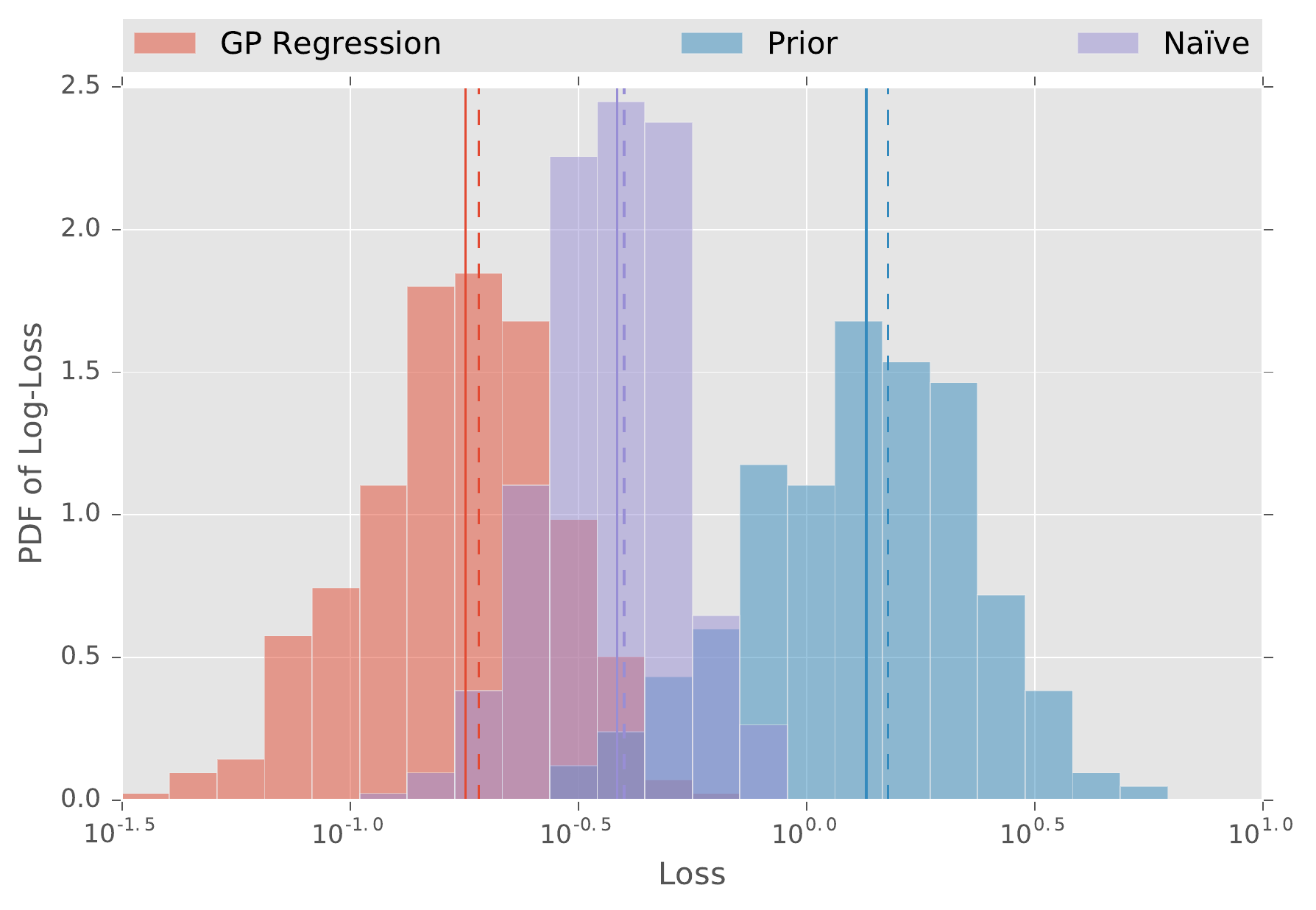}
	\end{center}
	\caption{\label{fig:gp-loss-hist}
		The performance of the Gaussian process estimator and naive estimator in relation to the prior loss. Plotted is a normalized histogram of the log-loss over 400 trials. The simulated experiment is that of $N=100$ (Left) and $N=1000$ (Right) single-shot repetitions of each of the 25 control sequences described in the text. Both the median loss (solid line) and the mean loss / Bayes risk (dashed line) are shown to guide the eye.
		The prior for the Gaussian process estimator is taken to be that shown in \autoref{fig:example-prior}, and the true spectra are sampled from the prior.
	}
\end{figure*}

We start with the true spectrum randomly selected by sampling a Gaussian Process, as exemplified in \autoref{fig:example-prior}. In this case, the hyperparameterized estimator is not useful\footnote{The hyperparameters are either not defined or have equivalent dimension to the GP estimator, in which case they would serve only to approximate something that can be analytically calculated.}, so we only compare the GP estimator with the naive estimator. As a point of reference, we can also treat the prior mean function as an estimator and calculate its loss. This is equivalent to the earlier comparison in \autoref{fig:example-gp-est}, but we average the results of many trials. The result of 400 trials is shown in \autoref{fig:gp-loss-hist}. As expected, the posterior loss is lower than the prior loss, indicating that the algorithm is learning. The naive loss does a respectable job as well, but is convincingly beaten by the GP estimator---especially given the fact that GP estimator comes with all the added benefits of the Bayesian methodology discussed above. 

\begin{figure*}
	\begin{center}
		\includegraphics[width=0.48\textwidth]{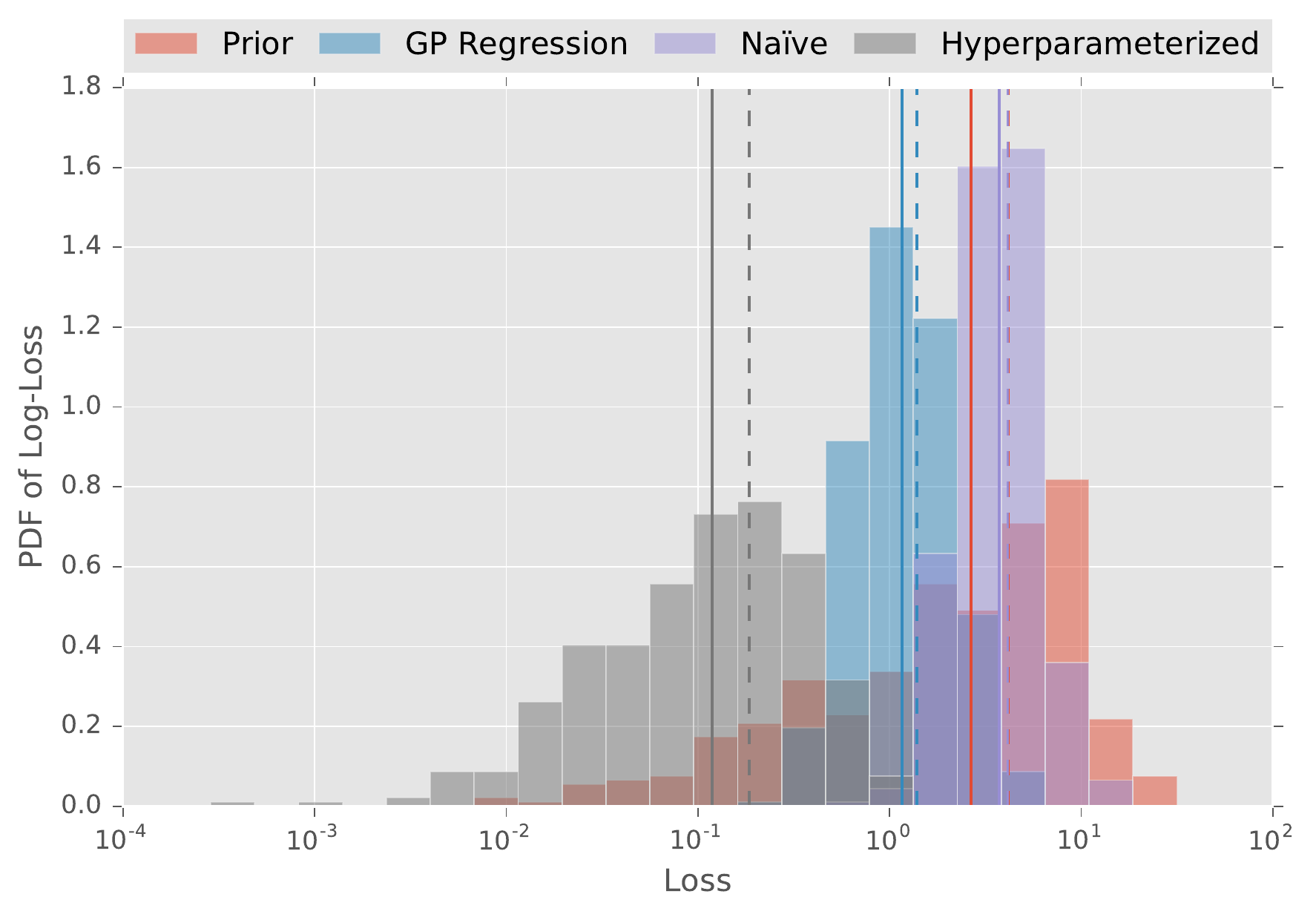}
		\includegraphics[width=0.48\textwidth]{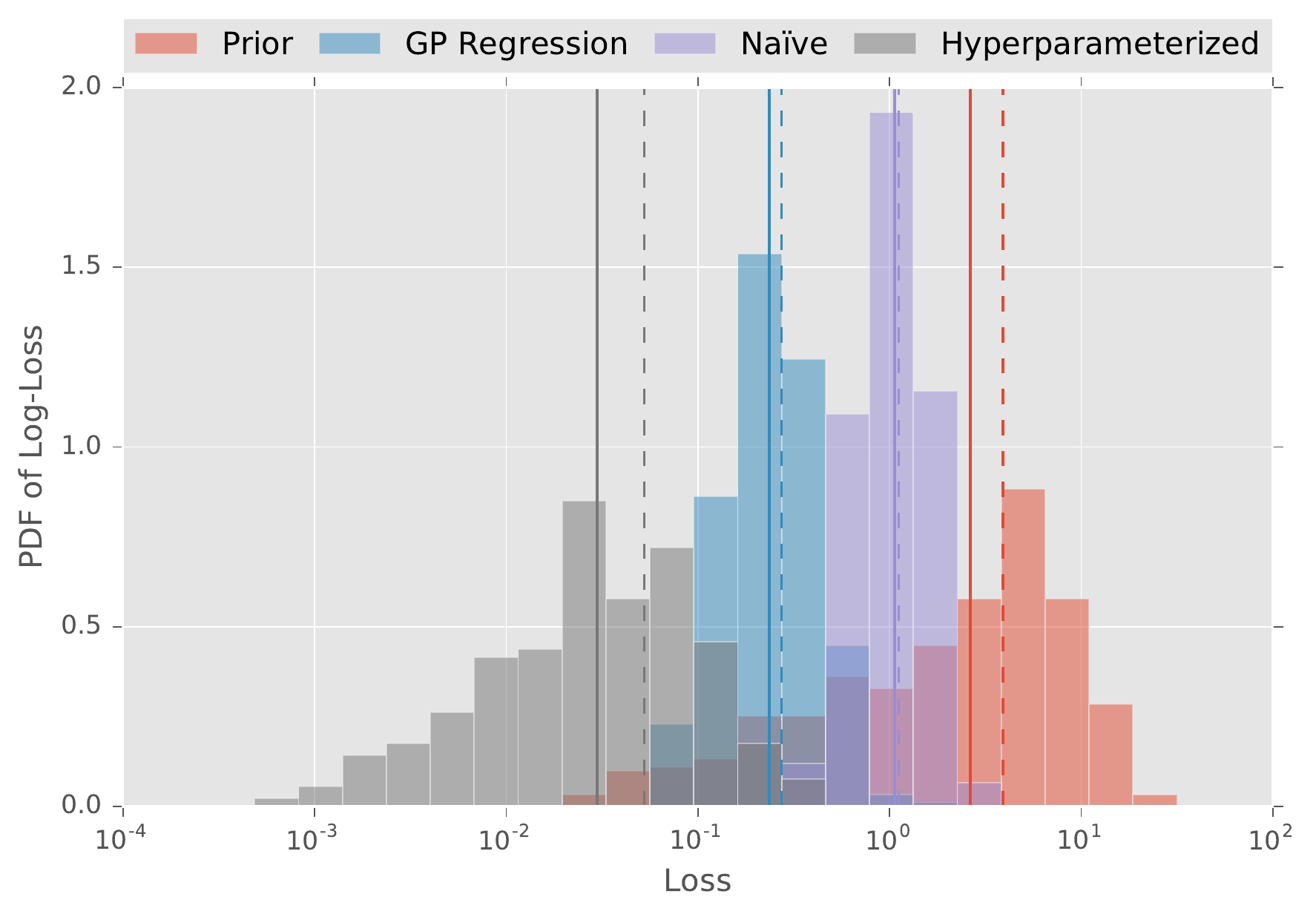}
	\end{center}
	\caption{\label{fig:one-on-f-loss-hist} 
		The performance of the hyperparameterized estimator, Gaussian Process estimator and naive estimator in relation to the prior loss. Plotted is a normalized histogram of the log-loss over 400 trials. The simulated experiment is that of $N=100$ (Left) and $N=1000$ (Right) single-shot repetitions of each of the 25 control sequences described in the text. Both the median loss (solid line) and the mean loss / Bayes risk (dashed line) are shown to guide the eye.
		For the prior on the hyperparemterized model, we use the hierarchal model \autoref{eq:one-on-f-cond}.
		For the Gaussian process estimator, we take the prior to have a mean function given by the $\mu(\omega) = \expect_{\alpha,A,c}[A / (\omega^{\alpha} + c)]$, where the expectation is over the prior for the hyperparameterized model.
		The covariance of the Gaussian process mean is taken to be the same kernel as that in \autoref{fig:example-prior}.
		Finally, the true spectra are drawn from the hyperparameterized prior.
	}
\end{figure*}

\begin{figure}
	\begin{center}
		\includegraphics[width=0.95\columnwidth]{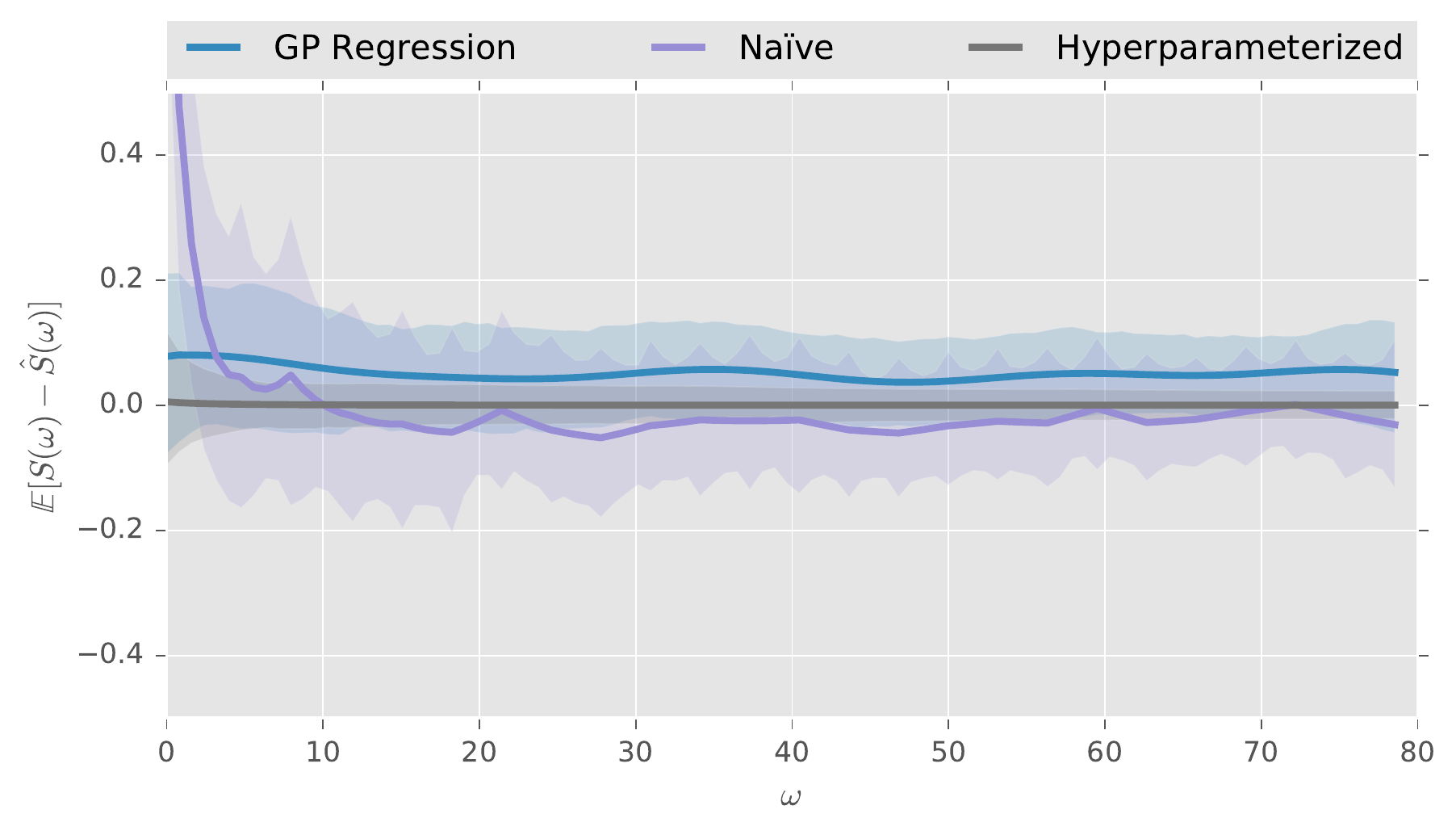}
	\end{center}
	\caption{\label{fig:one-on-f-residuals}
		The mean bias of the hyperparameterized estimator, Gaussian Process estimator and naive estimator as a function of $\omega$. The simulated experiment is that of 100 single-shot repetitions of each of the 25 control sequences described in the text. The solid lines are the mean performance over 400 trials while the shaded area indicates the range from the $25\%$ to the $75\%$ quantile.
	}
\end{figure}

In the case of the $1/f^\alpha$ model we compare all estimators. The results are presented in \autoref{fig:one-on-f-loss-hist} using another 400 trials. Again, we see that each estimator demonstrate genuine learning by reducing the loss over the prior. The GP still outperforms the naive estimator, but here the hyperparameterized estimator shines, demonstrating that knowledge of the additional structure is immensely beneficial for learning. We also plot the bias of each estimator in \autoref{fig:one-on-f-residuals}. The hyperparameterized estimator is extremely robust, reducing both the variance and bias over its competitors. By contrast, the GP estimator and naive estimator are reliably biased, and in opposite directions. We do not yet have a strong theoretical explanation for this behaviour.

\begin{figure}
	\begin{center}
		\includegraphics[width=0.95\columnwidth]{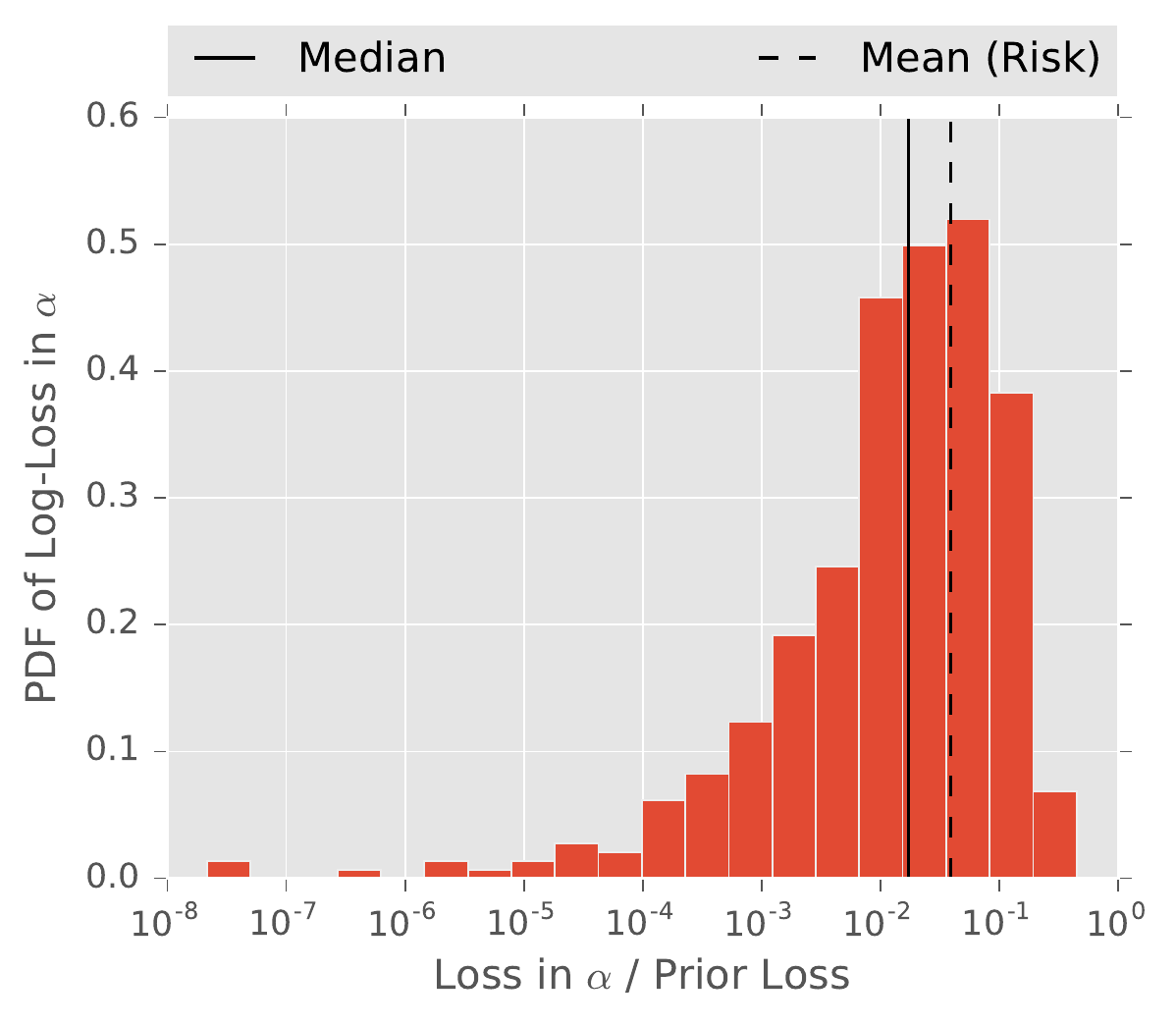}
	\end{center}
	\caption{\label{fig:hyperparm-loss}
		The performance of the hyperparameterized estimator, evaluated in terms of the
		hyperparameter $\alpha$. Plotted is a histogram of the hyperparameter loss
		$L_\alpha(\alpha, \hat{\alpha}) \defeq (\alpha - \hat\alpha)^2$ over 400 trials,
		normalized by the loss of the initial prior.
		The simulated experiment is that of 100 single-shot repetitions of each of the 25 control sequences described in the text. Both the median (solid line) and mean (dashed line) are shown to guide the eye.
	}
\end{figure}

When performing parametric estimation, as is implied by our hyperparameterized estimator, it is more common to define loss functions on the parameters themselves. The standard loss function is the squared error:
\begin{equation}
	L(\vec{\theta},\hat{\vec\theta}) = \|\vec{\theta}-\hat{\vec\theta}\|^2.
\end{equation}
As this loss function is not defined for the naive or GP estimator, we only report the results for the hyperparameterized estimator. This appears in \autoref{fig:hyperparm-loss}. Since we are only considering a single parametric estimator, in \autoref{fig:hyperparm-loss} we plot the ratio of the posterior loss to the prior loss of the hyperparameterized estimator.
This demonstrates that the parameter of interest, $\alpha$, can be learned to two orders of magnitude better accuracy than the prior with no more than 2,500 single-shot measurements.

%=============================================================================
\section{Discussion}
\label{sec:discussion}
%=============================================================================

In this work, we formulated the noise spectroscopy problem in the language of statistical estimation theory. This allows us to provide a robust and principled solution to the problem using Bayesian analysis. Considering \autoref{fig:cartoon} again, we have demonstrated two separate numerical solutions suited to two different regimes in the the continuum of possibilities: the big data limit and the low-data/low-dimension limit. 

In the large-data limit, we can effectively linearize the model and use a non-parametric approach to capture a broad class of spectra. The resultant spectra are not by eye different from a naive data fitting procedure in many cases. However, for a minimal amount of added computation, our approach gives a statistically rigorous accounting of the error bars in the reported spectra.

In the low-data limit we use the exact statistical model. In this case, estimation---or, \emph{learning}---is aided by use of prior knowledge on the model which reduces its dimension. The sequential Monte Carlo method is then applied to the resultant parameter estimation problem, which allows accurate inference even for highly non-linear models in the low data setting. 

In sum, we have treated what we consider the core inference problem in noise spectroscopy in order to provide the cleanest demonstration of our algorithms. Within our approach, however, there is no limit to the model complexity that can be treated---requiring only more computational resources. We comment briefly on some generalization that straightforwardly build on these methods. For instance, if one expects the power spectrum to have delta-like peaks---as would be the case if the probe was coupled to a finite number of harmonic oscillators, for example---it is possible to use the position and height of the peaks as hyperparameters in our routines. In an analogous way, our methods can be used for \emph{model selection}---that is, to discriminate between various proposed models for our environment. For instance, making contact with our previous example, one can determine how many oscillators are coupled to our probe. More importantly, the statistical methods developed here are not constrained to the spectroscopy scenario we considered: Gaussian, zero mean noise. More general spectroscopy protocols are based on inverting multi-dimensional integrals of the form   
$$ \int d\omega_1 \cdots d\omega_m F^{(m)}(\omega_1,\cdots,\omega_m,T) S^{(m)}(\omega_1, \cdots, \omega_m),$$
where $S^{(m)}(\omega_1, \cdots, \omega_m)$ is the $m$th order polyspectra and $F^{(m)}(\omega_1,\cdots,\omega_m,T)$ an $m$th generalized filter function \cite{norris_qubit_2016,paz-silva_multiqubit_2017}, essentially an $m$-th dimensional time ordered Fourier transform of the product $y(t_1) \cdots y(t_m)$. Estimating $S^{(m)}(\omega_1, \cdots, \omega_m)$ given our ability to manipulate $F^{(m)}(\omega_1,\cdots,\omega_m,T)$ is then a generalization of our current methods to higher dimensional integrals. 

Finally, we note that our inferential algorithm can easily be embedded into control software for online (real-time) estimation and, more interestingly, closed-loop adaptive control. These are exciting possibilities not currently offered without significant modification by traditional approaches.

%=============================================================================
% END MATTER
%=============================================================================

\begin{acknowledgments}
	
	CF was supported by the Australian Research Council Grant No. DE170100421. 
	CG was supported by the Australian Research Council via EQuS project number CE11001013, and 
    by the US Army Research Office grant numbers W911NF-14-1-0098. 
			GPS was supported by the Australian Research Council Grant No. DE170100088 and by a Griffith University Postdoctoral Fellowship. The ARC Centre of Excellence Grant No. CE110001027 (CQC2T) supported the research contributions of GPS and HMW.
\end{acknowledgments}

\onecolumngrid
\nocite{apsrev41Control}
\bibliographystyle{apsrev4-1}
\bibliography{apsrev-control,specdens}

\end{document}